# The transformation matrices (distortion, orientation, correspondence), their continuous forms, and their variants

Authors

**Cyril Cayron**[a]*

[a]Laboratory of ThermoMechanical Metallurgy (LMTM), PX Group Chair, Ecole Polytechnique Fédérale de Lausanne (EPFL), Rue de la Maladière, 71b, Neuchâtel, 2000, Switzerland

Correspondence email: cyril.cayron@epfl.ch

**Synopsis** Three transformation matrices (distortion, orientation, and correspondence) define the crystallography of displacive phase transformations. The paper explains how to calculate them and their variants, and why they should be distinguished.

**Abstract** The crystallography of displacive phase transformations can be described with three types of matrices: the lattice distortion matrix, the orientation relationship matrix, and the correspondence matrix. The paper gives some formula to express them in crystallographic bases, orthonormal bases, and reciprocal bases, and it explains how to deduce the matrices of inverse transformation. In the case of hard-sphere assumption, a continuous form of distortion matrix can be determined, and its derivative is identified to the velocity gradient used in continuum mechanics. The distortion, the orientation and the correspondence variants are determined by coset decomposition with intersection groups that depend on the point groups of the phases and on the type of transformation matrix. The stretch variants required in the phenomenological theory of martensitic transformation should be distinguished from the correspondence variants. The orientation variants and the correspondence variants are also different; they are defined from the geometric symmetries and algebraic symmetries, respectively. The concept of orientation (ir)reversibility during thermal cycling is briefly and partially treated by generalizing the orientation variants with *n*-cosets and graphs. Some simple examples are given to show that there is no general relation between the numbers of distortion, orientation and correspondence variants, and to illustrate the concept of orientation variants formed by thermal cycling.

## 1. Introduction

### 1.1. The transformation matrices

Martensitic phase transformation was first identified in steels more than a century ago. The transformation implies collective displacements of atoms (it is displacive); the parent austenite phase and the daughter martensite phase are linked by an orientation relationship (OR); and it often generates complex and intricate microstructures made of laths, plates or lenticles. The phenomenological theory of martensitic crystallography (PTMC) aims at explaining these features; it is based on linear algebra and three important matrices that we simply call here *transformation* matrices: the *distortion* matrix **F**, the *orientation* matrix **T**, and the *correspondence* matrix **C**. The matrix **F** tells how the crystallographic basis of the parent phase is distorted, the matrix **T** is a coordinate transformation matrix from the parent crystallographic basis to the daughter crystallographic basis (it encodes the orientation relationship), and the matrix **C** tells in which daughter crystallographic directions the directions of the parent crystallographic basis are transformed. These three matrices are used in PTMC to predict the habit planes and some variant pairing/grouping characteristics, as detailed in Appendix A1.

### 1.2. The variants

PTMC algorithms use mainly one type of variants, the stretch variants $\mathbf{U}_i$; the orientation variants $\mathbf{T}_i$ and the distortion variants $\mathbf{F}_i = \mathbf{Q}_i \mathbf{U}_i$ are outputs. Since in Bain's model of fcc-bcc martensite transformation there are thee stretch variants and thee correspondence variants, confusion may exist between stretch and correspondence variants. For example the variants $\mathbf{U}_i$ are called "correspondence variants" by Bhattacharya (2003), whereas they are, strictly speaking, stretch variants. In shape memory alloys (SMA) and ferroelectrics there is often a group-subgroup relation noted $\mathbb{G}^\alpha \leq \mathbb{G}^\gamma$ between the daughter phase $\alpha$ and the parent phase $\gamma$, with $\mathbb{G}^\alpha$ and $\mathbb{G}^\gamma$ the point groups of the phases. In this specific case, any matrix $\mathbf{U}_j = \mathbf{R}.\mathbf{U}_i.\mathbf{R}^{-1}$ where **R** belongs to $\mathbb{G}^\gamma$ but not to $\mathbb{G}^\alpha$ is a variant different from $\mathbf{U}_i$ (Bhattacharya, 2003). The number of stretch variants $N$ is simply the order $\mathbb{G}^\gamma$ divided by the order of $\mathbb{G}^\alpha$ i.e. $N = \frac{|\mathbb{G}^\gamma|}{|\mathbb{G}^\alpha|}$. The orientation variants are defined slightly differently; they are cosets of $\mathbb{G}^\alpha$ in $\mathbb{G}^\gamma$, but their number is also $\frac{|\mathbb{G}^\gamma|}{|\mathbb{G}^\alpha|}$, as shown by Janovec (1972, 1976). Should we conclude that when $\mathbb{G}^\alpha \leq \mathbb{G}^\gamma$, the stretch variants, the correspondence variants, and the orientation variants are always identical, or that, at least, their numbers are always equal?

In the absence of group-subgroup relation, the orientation variants are defined by cosets of $\mathbb{H}^\gamma$ in $\mathbb{G}^\gamma$, where $\mathbb{H}^\gamma \leq \mathbb{G}^\gamma$ is called "intersection group"; it is made of the symmetries that are common to both the parent and daughter phases (see for example Portier & Gratias, 1982; Dahmen, 1987; Cayron, 2006). The intersection group was introduced in metallurgy by Cahn & Kalonji (1981). It could be believed that the orientation variants form a group, but that is not true in general; actually, they have a groupoid structure. A groupoid can be understood as a generalized group whose structure that takes into account the local and global symmetries (see details and references in Cayron, 2006). One can also believe that there are as many orientation variants as distortion variants; but this not always true, as already illustrated with fcc-hcp transformation by Cayron (2016) and by Gao *et al.* (2016). Is there at least inequality relations between the numbers of orientation variants, stretch variants, distortion variants and correspondence variants?

**1.3. The orientation variants formed during thermal cycling**

The reversibility/irreversibility of martensitic alloys depends on many parameters. From a mechanical point of view, the compatibility between the austenite matrix and the martensite variants (or between the martensite variants themselves) plays a key role deeply treated in the modern versions of PTMC. The defects accumulated during thermal cycling are identified to elements of a group called "global group", which combines the symmetries and the lattice invariant shears (LIS). However, for reasons explained in Appendix A2, we prefer investigating another facet of reversibility/irreversibility that is only linked to the orientations, independently of any mechanical compatibility criterion. We note the orientation variants created by a series of *n* thermal cycles by $\gamma^1 \rightarrow \{\alpha^1\} \rightarrow \{\gamma^2\} \rightarrow \{\alpha^2\} \rightarrow \ldots \rightarrow \{\gamma^n\} \rightarrow \{\alpha^n\}$. Orientation reversibility is obtained when no new orientations of γ are created after a finite number of cycles, i.e. $\exists n \in \mathbb{N}, \{\alpha^n\} \subseteq \{\alpha^k\}, k \in [1, n-1]$. If the set of orientation variants $\{\gamma^2\}$ is reduced to a unique element which is the orientation of $\gamma^1$, the reversibility is obtained from the first cycle. It is often stated that this condition is satisfied when there is a group-subgroup relation, but it is very important to keep in mind that this relation means that the symmetries elements of the daughter phase should strictly coincide with those of the parent phase, and this condition depends on the OR. A transformation between a cubic γ phase and tetragonal α phase (with $a_\alpha = b_\alpha \neq c_\alpha$) with edge/edge <100>$_\gamma$ // <100>$_\alpha$ OR generates after one cycle the same orientation as the one of the initial parent γ crystal, and thus always comes back to this orientation by thermal cycling. However, a cubic γ - tetragonal α transformation with and $\frac{c_\alpha}{a_\alpha}$ irrational and with <110>$_\gamma$ // <101>$_\alpha$ OR generates by thermal cycling an infinite number of orientations. Therefore, it is important to mathematically specify the type of group-subgroup relation and the type of variants to which it applies.

**1.4. Objectives**

The aim of the paper is to give a coherent mathematical framework for the transformation matrices (correspondence, orientation, distortion) and for their variants. The matrices will be defined, and the methods to calculate them will be detailed. A continuous form of distortion matrix will be also introduced. It will be shown with geometric arguments that its multiplicative derivative is proportional to the velocity gradient of continuum mechanics. Continuous distortion offers new possible criteria beyond Schmid's law to explain the formation of martensite under stress. Formulae that give the correspondence, orientation, and distortion matrices of the reverse transformation as functions of the matrices of direct transformation will be presented. The orientation variants will be determined with the geometric symmetries, and the correspondence variants with the algebraic symmetries. The concept of group-subgroup relation will be specified according to these two types of symmetries. It will be shown that there are no general equalities or inequalities between the numbers of orientation, distortion and correspondence variants. Inequalities exist only for specific cases, such as with transformations implying a simple correspondence and an orientation group-subgroup relation. The orientation variants of direct and reverse transformations will be used to build the orientation graphs formed by thermal cycling and to discuss the conditions of orientation reversibility. Many 2D examples will be given to familiarize the reader with the different types of variants. We hope that the self-consistency of the paper will help clarifying the concept of transformation variants. It may be also useful in the future to continue building the bridge between crystallography and mechanics for phase transformations. The notation, described in Appendix B, may appear overloaded, but it is designed to respect the head-tail (target-source) composition rule.

## 2. Introduction to the transformation matrices

### 2.1. Distortion matrices

During displacive transformations the lattice of a parent crystal ($\gamma$) is distorted into the lattice of a daughter crystal ($\alpha$). In the case of deformation twinning, the parent and daughter phases are equal but the distortion restores the lattice into a new orientation. The distortion is assumed to be linear; it takes the form of an active matrix $\mathbf{F}^\gamma$. Any direction $\boldsymbol{u}$ is transformed by the distortion into a new direction $\boldsymbol{u}' = \mathbf{F}^\gamma \boldsymbol{u}$. The distortion matrix can be expressed in the usual crystallographic basis of the parent phase $\mathcal{B}_c^\gamma$, it is then noted $\mathbf{F}_c^\gamma$ and is given by $\mathbf{F}_c^\gamma = [\mathcal{B}_c^\gamma \rightarrow \mathcal{B}_c^{\gamma\prime}] = (\mathbf{a}^{\gamma\prime}, \mathbf{b}^{\gamma\prime}, \mathbf{c}^{\gamma\prime})$, writing the coordinates of the three vectors $\mathbf{a}^{\gamma\prime}, \mathbf{b}^{\gamma\prime}, \mathbf{c}^{\gamma\prime}$ in column in the basis $\mathcal{B}_c^\gamma$. One can also choose an orthonormal basis $\mathcal{B}_\# = (\boldsymbol{x}, \boldsymbol{y}, \boldsymbol{z})$ linked to the crystallographic basis $\mathcal{B}_c^\gamma$ by the structure tensor $\mathcal{S}^\gamma = [\mathcal{B}_\#^\gamma \rightarrow \mathcal{B}_c^\gamma]$ defined in equation (B13); the distortion in this basis is then noted $\mathbf{F}_\#^\gamma$. It is often easier to do the calculation in $\mathcal{B}_\#^\gamma$, and then coming back to the crystallographic basis by using equation (B6), $\mathbf{F}_c^\gamma = \mathcal{S}^{\gamma-1} \mathbf{F}_\#^\gamma \mathcal{S}^\gamma$.

In continuum mechanics, one would say that a material point $X$ follows a trajectory given by the positions $\boldsymbol{x} = \mathbf{F}.\boldsymbol{X}$, which implies that $d\boldsymbol{x} = \mathbf{F}.d\boldsymbol{X}$. The distortion matrix is thus $\mathbf{F} = \frac{d\boldsymbol{x}}{d\boldsymbol{X}} = (\nabla_X \boldsymbol{x})^\mathrm{t}$, the deformation gradient tensor (Bhattacharya, 2003).

## 2.1.1. Stretch matrices

One could think that polar decomposition can be directly applied to $\mathbf{F}_c^\gamma$ such that $\mathbf{F}_c^\gamma = \mathbf{R}_c^\gamma \mathbf{V}_c^\gamma$, where $\mathbf{V}_c^\gamma$ is a symmetric matrix in the crystallographic basis $\mathcal{B}_c^\gamma$. This is acceptable, but one should then keep in mind that then $\mathbf{V}_c^\gamma$ is not a stretch as we generally understand it, i.e. extensions or contractions along three perpendicular vectors. For example, a diagonal matrix written in a hexagonal basis means extension/contraction of the vectors of the basis, and these vectors are not perpendicular. If one wants to extract the "usual" stretch matrix from $\mathbf{F}_c^\gamma$, it is preferable to use $\mathbf{F}_\#^\gamma$. As the distortion matrix $\mathbf{F}_\#^\gamma$ is expressed in the orthonormal basis $\mathcal{B}_\#^\gamma$ it can be decomposed into

$$\mathbf{F}_\#^\gamma = \mathbf{Q}_\#^\gamma \mathbf{U}_\#^\gamma \qquad (1)$$

where $\mathbf{Q}_\#^\gamma$ is an orthogonal matrix and $\mathbf{U}_\#^\gamma$ is a symmetric matrix given by $\left(\mathbf{U}_\#^\gamma\right)^t \mathbf{U}_\#^\gamma = \left(\mathbf{U}_\#^\gamma\right)^2 = \left(\mathbf{F}_\#^\gamma\right)^t \mathbf{F}_\#^\gamma$. The matrix $\mathbf{U}_\#^\gamma$ can thus be written in another orthonormal basis as a diagonal matrix made of the square roots of its eigenvalues. Thus, $\mathbf{U}_\#^\gamma$ is a stretch matrix, sometimes called "Bain distortion" in tribute to the work of Bain (1924) on fcc-bcc transformations. Polar decomposition is an important tool in PTMC (Bhadeshia, 1987; Bhattacharya, 2003). Equation (1) can be then written in the crystallographic basis $\mathcal{B}_c^\gamma$ by

$$\mathbf{F}_c^\gamma = \left(\mathcal{S}^{\gamma-1}\mathbf{Q}_\#^\gamma \mathcal{S}^\gamma\right)\left(\mathcal{S}^{\gamma-1}\mathbf{U}_\#^\gamma \mathcal{S}^\gamma\right) = \mathbf{Q}_c^\gamma \mathbf{U}_c^\gamma \qquad (2)$$

Here, $\mathbf{U}_c^\gamma = \mathcal{S}^{\gamma-1}\mathbf{U}_\#^\gamma \mathcal{S}^\gamma$ expresses the same "usual" stretch as $\mathbf{U}_\#^\gamma$, but is not anymore necessarily a symmetric matrix because $\mathcal{S}^\gamma$ is not necessarily an orthogonal matrix. This shows that classical polar decomposition in a non-orthonormal basis does not necessarily result in a symmetric matrix.

The matrix $\mathbf{U}_c^\gamma$ can be obtained from $\mathbf{F}_c^\gamma$ directly working in $\mathcal{B}_c^\gamma$ by generalizing the polar decomposition to take into account the metrics; this is done with $\left(\mathbf{F}_c^\gamma\right)^t \mathcal{M}^\gamma \mathbf{F}_c^\gamma$ with $\mathcal{M}^\gamma$ the metric tensor of the γ phase. Indeed, by using equation (B16), we get $\left(\mathbf{F}_c^\gamma\right)^t \mathcal{M}^\gamma \mathbf{F}_c^\gamma = \left(\mathbf{U}_c^\gamma\right)^t \mathcal{M}^\gamma \mathbf{U}_c^\gamma = \mathcal{S}^{\gamma t}\left(\mathbf{U}_\#^\gamma\right)^2 \mathcal{S}^\gamma$; thus

$$\left(\mathbf{F}_c^\gamma\right)^t \mathcal{M}^\gamma \mathbf{F}_c^\gamma = \mathcal{M}^\gamma \left(\mathbf{U}_c^\gamma\right)^2 \qquad (3)$$

Polar decomposition permits to extract the stretch component of a distortion. The stretch contains the same information as the distortion matrix about the lattice strains because both matrices are related by a rotation. The change of free energy related to the transformation of a free (not constrained) single crystal is the same whether calculated with $\mathbf{U}_c^\gamma$ or with $\mathbf{F}_c^\gamma$. However, it is important to keep in mind that if the daughter product is formed inside a parent environment or as a wave that propagates at finite speed through a parent medium (Cayron, 2018), then the accommodation strains or the dominant wave vectors that could be calculated with $\mathbf{U}_c^\gamma$ are different from those calculated with $\mathbf{F}_c^\gamma$, and only the later is effective.

Two other points worth being mentioned: a) Left polar decomposition is in general different from a right polar decomposition because of the non-commutative character of matrix product. b) Polar decomposition and diagonalization are two methods that do not necessarily lead to the same result because the diagonalization is obtained in a basis that is not necessarily orthonormal. For example the one-step fcc-bcc distortion associated with Pitsch OR can be diagonalized with eigenvalues equal to 1, $\frac{\sqrt{8}}{3} \approx 0.943$ and $\frac{2}{\sqrt{3}} \approx 1.155$ (Cayron, 2013), whereas polar decomposition leads to the usual Bain stretch with values equal to $\sqrt{\frac{2}{3}} \approx 0.815, \frac{2}{\sqrt{3}}, \frac{2}{\sqrt{3}}$. The product of the eigenvalues is the same in both cases because the determinant of the distortion is the same and is equal to the fcc-bcc volume change, but the values are lower in the former case. The difference between the relations $\mathbf{F} = \mathbf{T}^{-1} \mathbf{W T}$ obtained by diagonalization with $\mathbf{W}$ is a diagonal matrix and $\mathbf{T}$ a coordinate transformation matrix, and $\mathbf{F} = \mathbf{Q} \mathbf{U}$ obtained by polar decomposition with $\mathbf{Q}$ a rotation and $\mathbf{U}$ a symmetric matrix, is fundamental and plays a key role in the spectral decomposition theorem.

**2.2. Orientation matrices**

The misorientation between the parent crystal γ and one of its variants α is given by the coordinate transformation matrix $\mathbf{T}_c^{\gamma \to \alpha} = [\mathcal{B}_c^\gamma \to \mathcal{B}_c^\alpha]$ (see section § B2 for the details). It is a passive matrix that changes the coordinates of a fixed vector $\boldsymbol{u}$ between the parent and daughter bases, $\boldsymbol{u}_{/\gamma} = \mathbf{T}^{\gamma \to \alpha} \boldsymbol{u}_{/\alpha}$. Sometimes the misorientation is given by the rotation that links the orthonormal bases of the parent and daughter crystals $\mathbf{R}_\#^{\gamma \to \alpha} = [\mathcal{B}_\#^\gamma \to \mathcal{B}_\#^\alpha]$. The misorientation and coordinate transformation matrices are linked by the relation $\mathbf{R}_\#^{\gamma \to \alpha} = \mathcal{S}^\gamma \mathbf{T}_c^{\gamma \to \alpha} \mathcal{S}^{\alpha-1}$.

All the matrices that encode the same misorientation as $\mathbf{T}_c^{\gamma \to \alpha}$ are obtained by multiplying $\mathbf{T}_c^{\gamma \to \alpha}$ at the right by the matrices $\boldsymbol{g}_i^\alpha \in \mathbb{G}^\alpha$ of internal symmetries of the daughter crystal:

$$\{\mathbf{T}_{ci}^{\gamma \to \alpha}\} = \{\mathbf{T}_c^{\gamma \to \alpha} \boldsymbol{g}_i^\alpha, \ \boldsymbol{g}_i^\alpha \in \mathbb{G}^\alpha\} \tag{4}$$

The set of equivalent rotations is thus $\{\mathbf{R}_\#^{\gamma \to \alpha}\} = \mathcal{S}^\gamma \{\mathbf{T}_c^{\gamma \to \alpha}\} \mathcal{S}^{\alpha-1}$. It is custom to choose the rotation with the lowest angle, called "disorientation". This choice has practical applications even if it remains arbitrary.

**2.3. Correspondence matrices**

The correspondence matrix $\mathbf{C}_c^{\alpha \to \gamma}$ gives the coordinates of the images by the distortion of the parent basis vectors, i.e. $\mathbf{a}^{\gamma\prime}, \mathbf{b}^{\gamma\prime}, \mathbf{c}^{\gamma\prime}$ expressed in the daughter basis $\mathcal{B}_c^\alpha$. Explicitly, $\mathbf{C}_c^{\alpha \to \gamma} = \left(\mathbf{a}_{/\mathcal{B}_c^\alpha}^{\gamma\prime}, \mathbf{b}_{/\mathcal{B}_c^\alpha}^{\gamma\prime}, \mathbf{c}_{/\mathcal{B}_c^\alpha}^{\gamma\prime}\right)$. The coordinates obey equation (B3) written as $\left(\mathbf{a}_{/\mathcal{B}_c^\alpha}^{\gamma\prime}, \mathbf{b}_{/\mathcal{B}_c^\alpha}^{\gamma\prime}, \mathbf{c}_{/\mathcal{B}_c^\alpha}^{\gamma\prime}\right) = \mathbf{T}_c^{\alpha \to \gamma} \left(\mathbf{a}_{/\mathcal{B}_c^\gamma}^{\gamma\prime}, \mathbf{b}_{/\mathcal{B}_c^\gamma}^{\gamma\prime}, \mathbf{c}_{/\mathcal{B}_c^\gamma}^{\gamma\prime}\right) = \mathbf{T}_c^{\alpha \to \gamma} \mathbf{F}_c^\gamma \left(\mathbf{a}_{/\mathcal{B}_c^\gamma}^{\gamma}, \mathbf{b}_{/\mathcal{B}_c^\gamma}^{\gamma}, \mathbf{c}_{/\mathcal{B}_c^\gamma}^{\gamma}\right) = \mathbf{T}_c^{\alpha \to \gamma} \mathbf{F}_c^\gamma$. Thus,

$$\mathbf{C}_c^{\alpha \to \gamma} = \mathbf{T}_c^{\alpha \to \gamma} \mathbf{F}_c^\gamma \tag{5}$$

As the correspondence is established for the three vectors of the parent basis, it is valid for the coordinates of any vector **x**:

$$\mathbf{x}'_{/\mathcal{B}_c^\gamma} = \mathbf{F}_c^\gamma \mathbf{x}_{/\mathcal{B}_c^\gamma} \Rightarrow \mathbf{x}'_{/\mathcal{B}_c^\alpha} = \mathbf{C}_c^{\alpha \rightarrow \gamma} \mathbf{x}_{/\mathcal{B}_c^\gamma} \tag{6}$$

The correspondence tells into which daughter directions the parent directions are transformed. It transforms all the rational vectors into other rational vectors; which is possible only if the matrix components are rational numbers, i.e. $\mathbf{C}_c^{\alpha \rightarrow \gamma} \in \text{GL}(3, \mathbb{Q})$, with $\text{GL}(3, \mathbb{Q})$ the general linear group of invertible matrices of dimension 3 defined on the field of rational numbers $\mathbb{Q}$.

The correspondence matrix $\mathbf{C}_c^{\alpha \rightarrow \gamma}$ associated with the distortion $\mathbf{F}_c^\gamma$ and the orientation $\mathbf{T}_c^{\gamma \rightarrow \alpha}$ is the same as that associated with the stretch distortion $\mathbf{U}_c^\gamma$ (linked to $\mathbf{F}_c^\gamma$ by polar decomposition $\mathbf{F}_c^\gamma = \mathbf{Q}_c^\gamma \mathbf{U}_c^\gamma$ given in equation (2)) and the orientation $(\mathbf{Q}_c^\gamma)^{-1} \mathbf{T}_c^{\gamma \rightarrow \alpha}$. Indeed, $\mathbf{C}_c^{\alpha \rightarrow \gamma}$ specifies the correspondence between the crystallographic directions (and chemical bonds), independently of any rigid-body rotation.

It is important to realize that there is no necessarily one-to-one relation between stretch and correspondence. Let us consider the 2D example of a square γ distorted into a square α such with a stretch matrix that is diagonal $\mathbf{U} = \begin{pmatrix} r & 0 \\ 0 & r \end{pmatrix}$ with $r = \frac{a_\gamma}{5a_\alpha}$, as shown in Figure 1. Two ORs are compatible with this distortion. The first one shown in Figure 1a is the OR $<1,0>_\gamma$ // $<5,0>_\alpha$, and the second one shown in Figure 1b is $<1,0>_\gamma$ // $<4,3>_\alpha$. The fact that the OR and the correspondence cannot be deduced automatically from the distortion comes when some directions or planes (parent or daughter) are equivalent by symmetry or by metrics (same length). The set of vectors that are "ambiguous" with a vector $u$ is defined by the equivalence class $\{v \in \mathbb{R}^3, \|v\| = \|u\|, v \notin \mathbb{G}.u\}$ with $\mathbb{G}$ the point group and $\|u\|$ the norm of $u$ defined in equation (B8). These ambiguities can be ignored most of the time, but they show that stretch and correspondence are different concepts that should be distinguished.

## 3. Construction of the transformation matrices with supercells

The simplest crystallographic transformation that can be imagined implies a one-to-one correspondence between the basis vectors $\mathbf{a}^\gamma \rightarrow \mathbf{a}^\alpha, \mathbf{b}^\gamma \rightarrow \mathbf{b}^\alpha, \mathbf{c}^\gamma \rightarrow \mathbf{c}^\alpha$, which means that $\mathbf{C}_c^{\alpha \rightarrow \gamma} = \mathbf{I}$. Some phase transformations are however more complex, as for fcc-bcc or fcc-hcp transformations. Whatever the complexity, the correspondence is always established between vectors of the Bravais lattices, i.e. vectors [u,v,w] with u,v,w integers or half-integers. Let us chose three of these vectors $\mathbf{u}^\gamma \rightarrow \mathbf{u}^{\gamma\prime} = \mathbf{u}^\alpha, \mathbf{v}^\gamma \rightarrow \mathbf{v}^{\gamma\prime} = \mathbf{v}^\alpha, \mathbf{w}^\gamma \rightarrow \mathbf{w}^{\gamma\prime} = \mathbf{w}^\alpha$ that are non-collinear and such that each of them has the lowest possible length. We build the supercell $\mathcal{B}_{super}^\gamma = (\mathbf{u}^\gamma, \mathbf{v}^\gamma, \mathbf{w}^\gamma)$. Its image by distortion is $\mathcal{B}_{super}^{\gamma\prime} = \mathcal{B}_{super}^\alpha = (\mathbf{u}^{\gamma\prime}, \mathbf{v}^{\gamma\prime}, \mathbf{w}^{\gamma\prime}) = (\mathbf{u}^\alpha, \mathbf{v}^\alpha, \mathbf{w}^\alpha)$. The correspondence, distortion and orientation can be defined only with this supercell. If the atoms inside the supercell do not follow the same trajectories as those at the corners of the cells, they are said to "shuffle". At each supercell, one introduce the matrices $\mathbf{B}^{\gamma}_{super} = [\mathcal{B}^{\gamma}_c \rightarrow \mathcal{B}^{\gamma}_{super}]$, $\mathbf{B}^{\gamma'}_{super} = [\mathcal{B}^{\gamma}_c \rightarrow \mathcal{B}^{\gamma'}_{super}]$ and $\mathbf{B}^{\alpha}_{super} = [\mathcal{B}^{\alpha}_c \rightarrow \mathcal{B}^{\alpha}_{super}]$. These three matrices are related to the distortion, orientation and correspondence matrices, as follows:

The *distortion matrix* is expressed in $\mathcal{B}^{\gamma}_{super}$ by $\mathbf{F}^{\gamma}_{super} = [\mathcal{B}^{\gamma}_{super} \rightarrow \mathcal{B}^{\gamma'}_{super}] = [\mathcal{B}^{\gamma}_{super} \rightarrow \mathcal{B}^{\gamma}_c][\mathcal{B}^{\gamma}_c \rightarrow \mathcal{B}^{\gamma'}_{super}] = (\mathbf{B}^{\gamma}_{super})^{-1}\mathbf{B}^{\gamma'}_{super}$. Written in the crystallographic basis $\mathcal{B}^{\gamma}_c$ with $\mathbf{F}^{\gamma}_c = \mathbf{B}^{\gamma}_{super}\mathbf{F}^{\gamma}_{super}(\mathbf{B}^{\gamma}_{super})^{-1}$, it gives

$$\mathbf{F}^{\gamma}_c = \mathbf{B}^{\gamma'}_{super}(\mathbf{B}^{\gamma}_{super})^{-1} \tag{7}$$

The *orientation matrix* is expressed in $\mathcal{B}^{\gamma}_{super}$ by $\mathbf{T}^{\gamma \rightarrow \alpha}_c = [\mathcal{B}^{\gamma}_c \rightarrow \mathcal{B}^{\alpha}_c] = [\mathcal{B}^{\gamma}_c \rightarrow \mathcal{B}^{\gamma'}_{super}][\mathcal{B}^{\gamma'}_{super} \rightarrow \mathcal{B}^{\alpha}_{super}][\mathcal{B}^{\alpha}_{super} \rightarrow \mathcal{B}^{\alpha}_c]$. Since $[\mathcal{B}^{\gamma'}_{super} \rightarrow \mathcal{B}^{\alpha}_{super}] = \mathbf{I}$, we get

$$\mathbf{T}^{\gamma \rightarrow \alpha}_c = \mathbf{B}^{\gamma'}_{super}(\mathbf{B}^{\alpha}_{super})^{-1} \tag{8}$$

The *correspondence matrix* is expressed in $\mathcal{B}^{\alpha}_{super}$ by $\mathbf{C}^{\alpha \rightarrow \gamma}_c = [\mathcal{B}^{\alpha}_c \rightarrow \mathcal{B}^{\gamma'}_c] = [\mathcal{B}^{\alpha}_c \rightarrow \mathcal{B}^{\gamma}_c][\mathcal{B}^{\gamma}_c \rightarrow \mathcal{B}^{\gamma'}_c]$, i.e. $\mathbf{C}^{\alpha \rightarrow \gamma}_c = \mathbf{T}^{\alpha \rightarrow \gamma}_c \mathbf{F}^{\gamma}_c$, as found in equation (5). According to the two previous equations,

$$\mathbf{C}^{\alpha \rightarrow \gamma}_c = \mathbf{B}^{\alpha}_{super}(\mathbf{B}^{\gamma}_{super})^{-1} \tag{9}$$

Since the matrices $\mathbf{B}^{\gamma}_{super}$ and $\mathbf{B}^{\alpha}_{super}$ are built on the crystallographic directions forming the supercell, their components are integers or half-integers; the correspondence matrix is thus always rational matrix, as already shown the previous section. In the case of a one-to-one correspondence between the basis vectors $\mathbf{B}^{\gamma}_{super} = \mathbf{B}^{\alpha}_{super} = \mathbf{I}$, and as expected $\mathbf{F}^{\gamma}_{super} = \mathbf{T}^{\gamma \rightarrow \alpha}_c$ and $\mathbf{C}^{\alpha \rightarrow \gamma}_c = \mathbf{I}$.

### 3.1. Reciprocal matrices

The distortion, orientation and correspondence matrices are defined for the crystallographic directions; i.e. the matrices are expressed in the direct space. The same operations can be determined for the crystallographic planes by writing the matrices in the reciprocal space. A plane $\boldsymbol{h}$ considered as a vector of the reciprocal space has its coordinates written in line, thus $\boldsymbol{h}^t$ written in column. The reciprocal distortion matrix $(\mathbf{F}^{\gamma}_c)^*$ is such that $(\boldsymbol{h}')^t = (\mathbf{F}^{\gamma}_c)^* \boldsymbol{h}^t$. Instead, one could have continued using $\boldsymbol{h}$ written in line, but in that case, the equation would have been $\boldsymbol{h}' = \boldsymbol{h}.(\mathbf{F}^{\gamma}_c)^{*\,t}$.

Any direction $\boldsymbol{u}$ of the plane $\boldsymbol{h}$ is such that the usual dot product $\boldsymbol{h}$ (in the reciprocal basis) by $\boldsymbol{u}$ (in the direct basis) is $\boldsymbol{h}.\boldsymbol{u} = 0$. After lattice distortion, the image of the plane is $\boldsymbol{h}'$ such that $\boldsymbol{h}'.\boldsymbol{u}' = 0$, i.e. $\boldsymbol{h}.(\mathbf{F}^{\gamma}_c)^{*\,t}\mathbf{F}^{\gamma}_c\boldsymbol{u} = 0$, which is verified for any vector $\boldsymbol{h}$ and $\boldsymbol{u}$ if and only if

$$(\mathbf{F}^{\gamma}_c)^* = (\mathbf{F}^{\gamma}_c)^{-t} \tag{10}$$

The same method could be used to show that

$$\left(\mathbf{T}_c^{\gamma\to\alpha}\right)^* = \left(\mathbf{T}_c^{\gamma\to\alpha}\right)^{-t} \text{ and } \left(\mathbf{C}_c^{\alpha\to\gamma}\right)^* = \left(\mathbf{C}_c^{\alpha\to\gamma}\right)^{-t} \qquad (11)$$

### 3.2. Matrices of reverse transformation

The orientation matrix of the direct transformation is $\mathbf{T}_c^{\gamma\to\alpha} = [\mathcal{B}_c^{\gamma} \to \mathcal{B}_c^{\alpha}]$ and that of the reverse transformation is $\mathbf{T}_c^{\alpha\to\gamma} = [\mathcal{B}_c^{\alpha} \to \mathcal{B}_c^{\gamma}]$. As $[\mathcal{B}_c^{\gamma} \to \mathcal{B}_c^{\alpha}][\mathcal{B}_c^{\alpha} \to \mathcal{B}_c^{\gamma}] = \mathbf{I}$, we get

$$\mathbf{T}_c^{\alpha\to\gamma} = \left(\mathbf{T}_c^{\gamma\to\alpha}\right)^{-1} \qquad (12)$$

Note that the index c means the crystallographic basis of the start lattice, i.e. $\mathcal{B}_c^{\gamma}$ for $\mathbf{T}_c^{\gamma\to\alpha}$ and $\mathcal{B}_c^{\alpha}$ for $\mathbf{T}_c^{\alpha\to\gamma}$. The correspondence matrix calculated with equation (9) is $\mathbf{C}_c^{\alpha\to\gamma} = \mathbf{B}_{super}^{\alpha}\left(\mathbf{B}_{super}^{\gamma}\right)^{-1}$ and that of the reverse transformation is $\mathbf{C}_c^{\gamma\to\alpha} = \mathbf{B}_{super}^{\gamma}\left(\mathbf{B}_{super}^{\alpha}\right)^{-1}$; consequently

$$\mathbf{C}_c^{\gamma\to\alpha} = \left(\mathbf{C}_c^{\alpha\to\gamma}\right)^{-1} \qquad (13)$$

As for the orientation matrix, the index c means $\mathcal{B}_c^{\alpha}$ for $\mathbf{C}_c^{\alpha\to\gamma}$ and $\mathcal{B}_c^{\gamma}$ for $\mathbf{C}_c^{\gamma\to\alpha}$.

The orientation and correspondence matrices of the reverse transformation are thus the inverse of the matrices of the direct transformation. This is not true for the distortion matrix. Indeed, the distortion matrix of the direct transformation in equation (5) is $\mathbf{F}_c^{\gamma} = \mathbf{T}_c^{\gamma\to\alpha}\mathbf{C}_c^{\alpha\to\gamma}$ and that of the reverse transformation is $\mathbf{F}_c^{\alpha} = \mathbf{T}_c^{\alpha\to\gamma}\mathbf{C}_c^{\gamma\to\alpha}$. They would be the inverse of the each other, $\mathbf{F}_c^{\alpha} = \left(\mathbf{F}_c^{\gamma}\right)^{-1}$, only if the product of the orientation matrix by the correspondence matrix were commutative, which is true only for specific cases, for example when $\mathbf{C}_c^{\alpha\to\gamma} = \mathbf{I}$. In the general case, the inversion relation does exist, but it appears when the matrices are written in the same basis. Indeed, the matrix $\mathbf{F}_c^{\alpha}$ is expressed in $\mathcal{B}_c^{\alpha}$, but when written in $\mathcal{B}_c^{\gamma}$ it becomes $\mathbf{F}_{c/\gamma}^{\alpha} = \mathbf{T}_c^{\gamma\to\alpha}\mathbf{F}_c^{\alpha}\mathbf{T}_c^{\alpha\to\gamma} = \mathbf{C}_c^{\gamma\to\alpha}\mathbf{T}_c^{\alpha\to\gamma} = \left(\mathbf{C}_c^{\alpha\to\gamma}\right)^{-1}\left(\mathbf{T}_c^{\gamma\to\alpha}\right)^{-1} = \left(\mathbf{T}_c^{\gamma\to\alpha}\mathbf{C}_c^{\alpha\to\gamma}\right)^{-1}$; thus

$$\mathbf{F}_{c/\gamma}^{\alpha} = \left(\mathbf{F}_c^{\gamma}\right)^{-1} \qquad (14)$$

The link between the $\mathbf{F}_c^{\alpha}$ and $\mathbf{F}_c^{\gamma}$ matrices expressed in their respective bases is

$$\mathbf{F}_c^{\alpha} = \mathbf{T}_c^{\alpha\to\gamma}\left(\mathbf{F}_c^{\gamma}\right)^{-1}\mathbf{T}_c^{\gamma\to\alpha} \qquad (15)$$

## 4. Continuous transformation matrices

### 4.1. Introduction to the angular parameter

In previous works (Cayron, 2015, 2016, 2017a,b, 2018), we assumed that the atoms in some simple metals behave as solid spheres during lattice distortion. This simplification, once associated with the knowledge of the parent-daughter orientation relationship, constrains the possible atomic displacements and the lattice distortion such that only one angular parameter becomes sufficient for their analytical determination. The distortion matrix expressed in $\mathcal{B}_c^{\gamma}$ appears as a function of the distortion angle θ, $\mathbf{F}_c^\gamma(\theta) = [\mathcal{B}_c^\gamma \to \mathcal{B}_c^\gamma(\theta)]$ with, let say, $\theta = \theta_1$ at the starting state, and $\theta = \theta_2$ when the transformation reaches completion. In the starting state, $\mathcal{B}_c^\gamma(\theta_1) = \mathcal{B}_c^\gamma$, the distortion matrix is $\mathbf{F}_c^\gamma(\theta_1) = \mathbf{I}$, and in the finishing state it is $\mathbf{F}_c^\gamma(\theta_2) = \mathbf{F}_c^\gamma$. There is no physical meaning for a continuous correspondence matrix $\mathbf{C}_c^{\gamma \to \alpha}$ or a continuous orientation matrix $\mathbf{T}_c^{\gamma \to \alpha}$ because these matrices take their significance only when the transformation is complete. However, we can assume that for any intermediate state αp (p for "precursor") that will become α when the transformation is finished, the correspondence matrix is already $\mathbf{C}_c^{\gamma \to \alpha p} = \mathbf{C}_c^{\gamma \to \alpha}$, and thus $\mathbf{T}_c^{\gamma \to \alpha p}(\theta) = \mathbf{F}_c^\gamma(\theta)$.

### 4.2. Continuous distortion matrix of reverse transformation

Let us consider the reverse transformation α→γ assuming that the direct transformation γ→α is already defined by $\mathbf{F}_c^\gamma(\theta)$. The same angular parameter θ can be chosen for the reverse transformation, but the start and finish angles should be exchanged, i.e. the distortion matrix in the start state becomes $\mathbf{F}_c^\alpha(\theta_2) = \mathbf{I}$, and in the finish state it is $\mathbf{F}_c^\alpha(\theta_1) = \mathbf{F}_c^\alpha$. The distortion matrix of any intermediate state is $\mathbf{F}_c^\alpha(\theta) = [\mathcal{B}_c^\alpha \to \mathcal{B}_c^\alpha(\theta)]$, with $\mathcal{B}_c^\alpha = \mathcal{B}_c^\alpha(\theta_2)$. Remember that $\mathbf{F}_c^\gamma(\theta)$ and $\mathbf{F}_c^\alpha(\theta)$, are expressed in their own basis, $\mathcal{B}_c^\gamma$ and $\mathcal{B}_c^\alpha$, respectively. Let us show how they are linked. The distortion $\mathbf{F}_c^\gamma(\theta)$ can be decomposed into two imaginary steps: (a) a complete transformation $\mathbf{F}_c^\gamma$ leading to the new phase α; then (b) a partial "come-back" step $\mathbf{F}_c^\alpha(\theta)$ in which the reverse transformation is stopped at the angle θ. As the matrices are active, they should be composed from the right to the left and be expressed in the same reference basis, here $\mathcal{B}_c^\gamma$; which gives $\mathbf{F}_c^\gamma(\theta) = \mathbf{F}_{c/\gamma}^\alpha(\theta)\, \mathbf{F}_c^\gamma$. By writing $\mathbf{F}_{c/\gamma}^\alpha(\theta)$ in the basis $\mathcal{B}_c^\alpha$ with the help of the coordinate transformation matrix $\mathbf{T}_c^{\gamma \to \alpha}$, and by using equation (5), it comes

$$\mathbf{F}_c^\gamma(\theta) = \mathbf{T}_c^{\gamma \to \alpha} \mathbf{F}_c^\alpha(\theta)\, \mathbf{C}_c^{\alpha \to \gamma} \tag{16}$$

One can check that $\mathbf{F}_c^\gamma(\theta_2) = \mathbf{F}_c^\gamma$ and $\mathbf{F}_c^\alpha(\theta_1) = \mathbf{F}_c^\alpha$. This approach was used by Cayron (2016) to calculate the bcc-fcc continuous distortion matrix from the fcc-bcc one.

### 4.3. Derivative of continuous distortion matrices

#### 4.3.1. Geometric introduction to multiplicative derivative

Our theoretical work on the crystallography of martensitic transformation for the last years is driven by our will to understand the continuous features observed in the pole figures of martensite in steels (see § A3). Different models of fcc-bcc transformation were proposed; the continuous features are qualitatively explained, but quantitative simulations based on rigorous mathematics are still missing. The main obstacle is the way to extract the "rotational" part of a distortion matrix. We have seen in § 2.1.1 that polar decomposition gives different results depending on choice of the decomposition (left or right). One way to tackle the problem is to work with infinitesimal distortions. For this aim we introduce $D\mathbf{F}$, the infinitesimal distortion matrix $\mathbf{F}$, locally defined in $\mathcal{B}(\theta)$ by $D\mathbf{F}(\theta)_{loc} = [\mathcal{B}(\theta) \rightarrow \mathcal{B}(\theta + d\theta)]$, as shown in Figure 2. This element can be decomposed into two imaginary steps $[\mathcal{B}(\theta) \rightarrow \mathcal{B}(\theta_s)] [\mathcal{B}(\theta_s) \rightarrow \mathcal{B}(\theta + d\theta)]$ with $\theta_s$ the angular parameter at the start state; thus

$$D\mathbf{F}(\theta)_{loc} = \mathbf{F}(\theta)^{-1} \mathbf{F}(\theta + d\theta). \tag{17}$$

If the basis of the starting state $\mathcal{B}(\theta_s)$ is used as an absolute basis,

$D\mathbf{F}(\theta) = [\mathcal{B}(\theta_s) \rightarrow \mathcal{B}(\theta)] D\mathbf{F}(\theta)_{loc} [\mathcal{B}(\theta) \rightarrow \mathcal{B}(\theta_s)] = \mathbf{F}(\theta + d\theta) \mathbf{F}(\theta)^{-1}$; which we simply write

$$D\mathbf{F}(\theta) = \mathbf{F}(\theta + d\theta) \mathbf{F}(\theta)^{-1} \tag{18}$$

One can also understand equation (18) in its active meaning, by writing that the images at $\theta$ and $\theta + d\theta$ of a fixed vector $\mathbf{u}_0$ are $\mathbf{u}(\theta) = \mathbf{F}(\theta)\mathbf{u}_0$ and $\mathbf{u}(\theta + d\theta) = \mathbf{F}(\theta + d\theta) \mathbf{u}_0$, thus $\mathbf{u}(\theta + d\theta) = \mathbf{F}(\theta + d\theta) \mathbf{F}(\theta)^{-1}\mathbf{u}(\theta) = D\mathbf{F}(\theta)\mathbf{u}(\theta)$. If $\mathbf{F}$ is a one-dimensional function $f$ of $\theta$, this infinitesimal form is $Df = \frac{f(\theta+d\theta)}{f(\theta)}$. Note that $D\mathbf{F}$ is different from $d\mathbf{F}(\theta) = \mathbf{F}(\theta + d\theta) - \mathbf{F}(\theta)$, which is the usual infinitesimal form of each of the nine components of $\mathbf{F}(\theta)$. In other words, $D\mathbf{F}$ and $d\mathbf{F}$ are respectively given by the ratio and difference of two infinitesimally close terms; they are close to 1 and 0, respectively. $D\mathbf{F}$ as a "multiplicative" infinitesimal and $d\mathbf{F}$ is an "additive" infinitesimal; they are linked together by

$$D\mathbf{F} = \mathbf{I} + d\mathbf{F}\, \mathbf{F}^{-1} \tag{19}$$

The integral of the additive infinitesimal and the integrals of the local and global multiplicative infinitesimals are also linked together by

$$\mathbf{F}(\theta) = \int_\theta d\mathbf{F} = \prod_{\theta \rightarrow} D\mathbf{F}_{loc} = \prod_{\theta \leftarrow} D\mathbf{F} \tag{20}$$

where $\int$ is the usual continuous integration symbol and $\prod$ is the continuous multiplicative integration symbol, with $\theta \rightarrow$ and $\theta \leftarrow$ meaning that the product should be made from the left to the right (passive way) and from the right to the left (active meaning), respectively. Equation (19) also permits to define the multiplicative derivative of $\mathbf{F}$ by

$$\frac{D\mathbf{F}}{D\theta} = \frac{D\mathbf{F} - \mathbf{I}}{d\theta} = \frac{d\mathbf{F}}{d\theta}\mathbf{F}^{-1} \tag{21}$$

The multiplicative derivation and integration are natural for matrices because they take into consideration the non-commutativity of the matrix product. Their use is unfortunately not widespread in physics, despite their early introduction by Volterra (1887) for sets of differential equations. The multiplicative derivation and integration, and many related formula, are detailed in a recent book (Slavik, 2007).

<u>Note</u>: One could believe that, as $\mathbf{F}(\theta)$ and $\mathbf{F}(\theta)^{-1}$ commute, then $\mathbf{F}(\theta + d\theta)$ and $\mathbf{F}(\theta)^{-1}$ also commute when $d\theta \rightarrow 0$; which would allow us to use equation (18) to write an equation that would link the two types of infinitesimal: $ln(D\mathbf{F}) = d(ln\mathbf{F})$. However, $\mathbf{F}(\theta + d\theta)$ and $\mathbf{F}(\theta)^{-1}$ do *not* commute in general. If it were the case, equation (21) would be written from two products $\frac{d\mathbf{F}}{d\theta}\mathbf{F}^{-1}$ and $\mathbf{F}^{-1}\frac{d\mathbf{F}}{d\theta}$ that should be equal, which is not true, as the reader can check with by the simple example $\mathbf{F}(\theta) = \begin{pmatrix} 1 & Cos(\theta) \\ 0 & Sin(\theta) \end{pmatrix}$. Deep and careful reading of mathematical literature would be required to use properly the logarithm and exponential functions of matrices of type $\mathbf{F}(\theta)$ in order to link the multiplicative and additive derivatives.

### 4.3.2. Possible applications of the continuous distortion matrices

One can recognise in equation (21) the velocity gradient $\mathbf{L} = \dot{\mathbf{F}}\,\mathbf{F}^{-1}$ defined in continuum mechanics in which $\dot{\mathbf{F}}$ is the usual derivative of $\mathbf{F}$ on time. More precisely,

$$\mathbf{L}(\theta) = \dot\theta\,\frac{D\mathbf{F}}{D\theta} \qquad (22)$$

Therefore, as in continuum mechanics, the rotational part of a continuous distortion matrix $\mathbf{F}(\theta)$ can be defined at each value θ by $Spin\mathbf{F}(\theta) = \frac{\mathbf{L}(\theta)-\mathbf{L}(\theta)^t}{2}$. We are currently working at introducing $Spin\mathbf{F}$ to explain the continuities observed in the pole figures of martensite.

The multiplicative derivative of the distortion matrix can be also used to propose crystallographic criteria that aim at predicting mechanical conditions in which the transformation can be triggered. It is usual in mechanics to assume that deformation twinning is a simple shear mechanism, and that twinning occurs when the crystallographic shear plane and direction coincide with the applied shear stress (Schmid's law). However, many experimental studies report some deformation twinning modes in magnesium with abnormal Schmid factors (Beyerlein *et al*., 2010), or martensite variant reorientations in SMAs that are not in agreement with simple shears (Alkan, Wu & Sehitoglu, 2017; Bucsek *et al*., 2018). We think that the inadequacy of Schmid's law for phase transformations comes from the fact that a continuous simple strain path is not compatible with realistic atomic interactions. Due to atomic steric effects, a simple shear (stress) does not induce a simple strain (deformation) (Cayron, 2018). In order to take into account this effect, we proposed that twins or martensite appear for positive mechanical interaction work W (Cayron, 2017a). The interaction work is given by the Frobenius inner product $W = \boldsymbol{\sigma}_{ij}\boldsymbol{\varepsilon}_{ij}$, i.e. by the addition of the term-by-term products along the indices *i,j* of $\boldsymbol{\sigma}$ the external applied stress tensor, and $\boldsymbol{\varepsilon}$ the deformation tensor calculated at the intermediate state $\theta_{\text{int}}$ where the volume change is maximum during lattice distortion $\boldsymbol{\varepsilon} = \mathbf{F}(\theta_{int}) - \mathbf{F}(\theta_s) = \mathbf{F}(\theta_{int}) - \mathbf{I}$. Another criterion based on the derivative of the distortion matrix could be also imagined, for example by introducing the angular power $\dot{W} = \boldsymbol{\sigma}_{ij}\,\dot{\boldsymbol{\varepsilon}}_{ij}$ at the starting state, with $\dot{\boldsymbol{\varepsilon}} = \frac{D\mathbf{F}}{D\theta}\bigg|_{\theta=\theta_s} = \frac{d\mathbf{F}}{d\theta}\bigg|_{\theta=\theta_s}$.

Let us illustrate these different possible criteria with the simple 2D case of α→α$_t$ deformation twinning of a Pmm phase shown in Figure 3a. If the atoms are considered as hard-spheres, the atomic displacements along a simple strain trajectory are impossible due to the steric effect. If the crystal is free, it is reasonable to assume that the atoms actually "roll" on each other, and that the lattice continuously switches from the starting state at $\beta_s = \frac{\pi}{3}$ to the finishing state at $\beta_f = \frac{2\pi}{3}$. It can be noted that the distortion implies an intermediate transitory cubic state at $\beta_{int} = \frac{\pi}{2}$ with a volume change allowed by the absence of external constraints. The distortion matrix can be calculated in the orthonormal basis of this intermediate cubic phase $\mathcal{B}_\#$. The supercell marked by the vectors $\mathcal{B}^\alpha_{super}(\beta) = (\mathbf{u}^\alpha, \mathbf{v}^\alpha)$ is expressed by the matrix $\mathbf{B}^\alpha_{super}(\beta) = [\mathcal{B}_\# \to \mathcal{B}^\alpha_{super}(\beta)] = \begin{pmatrix} 1 & Cos(\beta) \\ 0 & Sin(\beta) \end{pmatrix}$. Applying equation (7) in $\mathcal{B}_\#$ leads to $\mathbf{F}^\alpha_\#(\beta) = \mathbf{B}^\alpha_{super}(\beta)\left(\mathbf{B}^\alpha_{super}(\beta_s)\right)^{-1} = \begin{pmatrix} 1 & \frac{-1+2Cos(\beta)}{\sqrt{3}} \\ 0 & \frac{2Sin(\beta)}{\sqrt{3}} \end{pmatrix}$. The volume change during twinning, given by the determinant of the distortion, is shown in Figure 3b. The displacements of the atom located in $\frac{1}{2}\mathbf{v}^\alpha$ following a simple strain, a distortion at the maximum volume change, or the distortion derivative, are given by the vectors $\mathbf{d}^s_\# = \left(\mathbf{F}^\alpha_\#\left(\beta_f\right) - \mathbf{I}\right) \cdot \frac{1}{2}\mathbf{v}^\alpha_\#$, $\mathbf{d}^m_\# = \left(\mathbf{F}^\alpha_\#\left(\beta_{int}\right) - \mathbf{I}\right) \cdot \frac{1}{2}\mathbf{v}^\alpha_\#$, or $\dot{\mathbf{d}}_\# = \left(\frac{d\mathbf{F}_\#(\beta)}{d\beta}\bigg|_{\beta=\beta_s}\right) \cdot \frac{1}{2}\mathbf{v}^\alpha_\#$, respectively. Since $\mathbf{v}^\alpha_\# = [Cos(\beta_s), Sin(\beta_s)]^t = \left[\frac{1}{2}, \frac{\sqrt{3}}{2}\right]^t$, we get $\mathbf{d}^s_\# = \left[-\frac{1}{2}, 0\right]^t$, $\mathbf{d}^m_\# = \left[-\frac{1}{4}, \frac{1}{2} - \frac{\sqrt{3}}{4}\right]^t$, and $\dot{\mathbf{d}}_\# = \left[-\frac{\sqrt{3}}{4}, \frac{1}{4}\right]^t$. The three types of displacements are represented in Figure 3c. The displacement $\dot{\mathbf{d}}$ calculated with the derivative of the distortion matrix is perpendicular to $\frac{1}{2}\mathbf{v}^\alpha$, as expected for a displacement that is compatible with the hard-sphere assumption. The angular power criterion can be understood as a triggering criterion that only takes into account the steric effect of the atoms at the first instants of the distortion process. Baur *et al.* (2017b) shows that this criterion could explain variant selection of the martensite formed at the surface of Fe-Ni alloys during electropolishing. More experimental works on perfectly oriented single crystals are required to compare the predictions based on simple shear (and Schmid's law) with those based on maximum volume change or on the derivative of the angular distortion.

It was noticed in Figure 3 that it is impossible to continuously transform the crystal α into its twin α$_t$ without passing by a transient cubic state. One can thus wonder whether stress-induced reorientation of variants in SMAs can really be obtained by a continuous simple strain of type $\mathbf{P}_{21}$ (see § A1), or if the steric effect impedes this simple strain and necessarily implies a high symmetry transient state. In the case of NiTi, it would mean that stress-induced reorientation B19' (variant 2) → B19' (variant 1) would be actually a double-step mechanism B19' (variant 2) → B2 (parent)→ B19' (variant 1). Such an idea could have implication to explain the tension/compression asymmetry in SMA (Liu *et al.*, 1998); this asymmetry is indeed difficult to understand with Schmid's law because a simple strain that would induce B19' (variant 2) → B19' (variant 1) reorientation is the reverse of the simple strain of B19' (variant 1) → B19' (variant 2) reorientation. This is not the case with a criterion based on angular distortion. However, here again, more works are required to test this speculative hypothesis.

**5. Variants**

**5.1. Orientation variants**

The orientation variants only depend on the orientation relationship. They can be mathematically defined from the subgroup $\mathbb{H}_T^\gamma$ of the symmetries that are common to the parent and daughter crystals

$$\mathbb{H}_T^\gamma = \mathbb{G}^\gamma \cap \mathbf{T}_c^{\gamma\to\alpha}\mathbb{G}^\alpha \left(\mathbf{T}_c^{\gamma\to\alpha}\right)^{-1} \tag{23}$$

with $\mathbf{T}_c^{\gamma\to\alpha}$ the orientation matrix, and $\mathbb{G}^\gamma$ and $\mathbb{G}^\alpha$ the point groups of parent and daughter phases. The matrix $\mathbf{T}_c^{\gamma\to\alpha}$ is used to express in the parent basis the geometric symmetry elements of the daughter phase. Geometrically, the intersection group $\mathbb{H}_T^\gamma$ is made of the parent and daughter symmetry elements that are in coincidence. It will be shown in § 5.3 that another type of intersection group based on the correspondence matrix and algebraic symmetries can be also built.

The orientation variants are defined by the cosets $\alpha_i = g_i^\gamma \mathbb{H}_T^\gamma$ and their orientations are $\mathbf{T}^{\gamma\to\alpha_i} = g_i^\gamma \mathbf{T}^{\gamma\to\alpha}$ with $g_i^\gamma \in \alpha_i$ (matrices of the coset $\alpha_i$), as detailed in (Cayron, 2006). All the matrices that belong to $\mathbb{H}_T^\gamma$ point to the same orientation variant $\alpha_1$, all the matrices that belong to $g_2^\gamma \mathbb{H}_T^\gamma$ with $g_2^\gamma \notin \mathbb{H}_T^\gamma$ point to the orientation variant $\alpha_2$, etc. The number of orientation variants is thus

$$N_T^\alpha = \frac{|\mathbb{G}^\gamma|}{|\mathbb{H}_T^\gamma|} \tag{24}$$

Let us consider formula (24) for the reverse transformation α→γ, $N_T^\gamma = \frac{|\mathbb{G}^\alpha|}{|\mathbb{H}_T^\alpha|}$. As $\mathbb{H}_T^\alpha$ and $\mathbb{H}_T^\gamma$ are linked by the isomorphism $\mathbb{H}_T^\gamma = \mathbf{T}_c^{\gamma\to\alpha}\mathbb{H}_T^\alpha \left(\mathbf{T}_c^{\gamma\to\alpha}\right)^{-1}$, their orders are equal: $|\mathbb{H}_T^\alpha| = |\mathbb{H}_T^\gamma|$; which is expected because both sides of the equation give the number of common geometric symmetries. The number of variants of the direct transformation and the number of variants of the reverse transformation are linked by the relation (Cayron, 2006):

$$N_T^\alpha |\mathbb{G}^\alpha| = N_T^\gamma |\mathbb{G}^\gamma| \tag{25}$$

When there is an *orientation* group-subgroup relation between the daughter and parent phases $\mathbf{T}_c^{\gamma\to\alpha}\mathbb{G}^\alpha \left(\mathbf{T}_c^{\gamma\to\alpha}\right)^{-1} \leq \mathbb{G}^\gamma$, thus $\mathbb{H}_T^\gamma = \mathbf{T}_c^{\gamma\to\alpha}\mathbb{G}^\alpha \left(\mathbf{T}_c^{\gamma\to\alpha}\right)^{-1} \equiv \mathbb{G}^\alpha$, and $N_T^\alpha = \frac{|\mathbb{G}^\gamma|}{|\mathbb{G}^\alpha|}$. The number of γ orientations created by the reverse transformation of the α variants is $N_T^\gamma = \frac{|\mathbb{G}^\alpha|}{|\mathbb{H}_T^\alpha|} = 1$; thus, there is no new orientations created by cycles of transformation, the parent γ crystal always come back to its initial orientation.

## 5.2. The different types of distortion variants

There are two important kinds of distortion variants, one based on how the symmetries act on the distortion, and the other one on how the distortion acts on the symmetries. The former are simply called here *distortion* variants, and the latter *distorted-shape* variants. The stretch variants are derivative of distortion variants.

### 5.2.1. Distortion variants

The distortion variants can be defined only when the mechanism of transformation implies a lattice distortion, which is clearly the case for displacive transformation and for diffusion-limited displacive transformation (bainite, shape memory alloys). They are not relevant for usual precipitation, for which the variants are dictated only by the orientation relationship matrix (§ 5.1). The distortion variants can be introduced by considering all the distinct distortion matrices that can be calculated from the symmetries of the parent phase. In each equivalent basis $\mathcal{B}_{ci}^{\gamma} = g_i^{\gamma} \mathcal{B}_{c}^{\gamma}$ of the parent crystal, the distortion matrices are locally written as $\mathbf{F}_c^{\gamma}$. Once expressed in the reference basis $\mathcal{B}_{c}^{\gamma}$, they become $g_i^{\gamma} \mathbf{F}_c^{\gamma} (g_i^{\gamma})^{-1}$. The set of all possible distortion matrices is thus $\mathbb{O}^{\gamma} = \{g_i^{\gamma} \mathbf{F}_c^{\gamma} (g_i^{\gamma})^{-1}, g_i^{\gamma} \in \mathbb{G}^{\gamma}\}$. Algebraically, the group $\mathbb{G}^{\gamma}$ acts by conjugation on $\mathbf{F}_c^{\gamma}$, and $\mathbb{O}^{\gamma}$ is the orbit of the conjugacy action of $\mathbb{G}^{\gamma}$ on $\mathbf{F}_c^{\gamma}$. The stabilizer of $\mathbf{F}_c^{\gamma}$ is a subgroup $\mathbb{G}^{\gamma}$ constituted by the matrices $g_i^{\gamma}$ that leave $\mathbf{F}_c^{\gamma}$ invariant by the conjugacy action; it is

$$\mathbb{H}_F^{\gamma} = \{ g_i^{\gamma} \in \mathbb{G}^{\gamma}, \ g_i^{\gamma} \mathbf{F}_c^{\gamma} (g_i^{\gamma})^{-1} = \mathbf{F}_c^{\gamma}\} \tag{26}$$

The number of distinct conjugated matrices, i.e. the number of distortion variants, is the number of elements in the orbit $\mathbb{O}^{\gamma}$; it is given by the orbit-stabilizer theorem:

$$N_F^{\alpha} = |\mathbb{O}^{\gamma}| = \frac{|\mathbb{G}^{\gamma}|}{|\mathbb{H}_F^{\gamma}|} \tag{27}$$

The other way to figure out the distortion variants consists in changing equation (26) as

$$\mathbb{H}_F^{\gamma} = \{ g_i^{\gamma} \in \mathbb{G}^{\gamma}, \ (\mathbf{F}_c^{\gamma})^{-1} g_i^{\gamma} \mathbf{F}_c^{\gamma} = g_i^{\gamma}\} \tag{28}$$

This is the subgroup of symmetries elements left invariant by the distortion $\mathbf{F}_c^{\gamma}$. The number of distortion variants is then deduced directly from Lagrange; it is formula (27).

Now, if we consider the reverse transformation, $N_F^{\gamma} = \frac{|\mathbb{G}^{\alpha}|}{|\mathbb{H}_F^{\gamma}|}$. The stabilizer is $\mathbb{H}_F^{\alpha} = \{ g_i^{\alpha} \in \mathbb{G}^{\alpha}, \ g_i^{\alpha} \mathbf{F}_c^{\alpha} (g_i^{\alpha})^{-1} = \mathbf{F}_c^{\alpha}\}$. By using equation (14) written as $(\mathbf{F}_c^{\gamma})^{-1} = \mathbf{T}_c^{\gamma \to \alpha} \mathbf{F}_c^{\alpha} (\mathbf{T}_c^{\gamma \to \alpha})^{-1}$, one can check that $\mathbb{H}_F^{\gamma} = \mathbf{T}_c^{\gamma \to \alpha} \mathbb{H}_F^{\alpha} (\mathbf{T}_c^{\gamma \to \alpha})^{-1}$. This establishes an isomorphism between the two subgroups, thus $|\mathbb{H}_F^{\alpha}| = |\mathbb{H}_F^{\gamma}|$; which is expected because both sides of the equation give the number of symmetries invariant by the distortion. Consequently, formula (25) obtained for the orientation variants, also holds for the distortion variants, i.e.

$$N_{\text{F}}^{\alpha} |\mathbb{G}^{\alpha}| = N_{\text{F}}^{\gamma}|\mathbb{G}^{\gamma}| \qquad (29)$$

### 5.2.2. Distorted-shape variants

Another type of distortion variant now based on the crystal shape and its symmetries can be imagined. A symmetry of the parent crystal expressed by $g_i^{\gamma} \in \mathbb{G}^{\gamma}$ in the basis $\mathcal{B}_c^{\gamma}$ continues to be a symmetry after distortion if it is expressed in the basis $\mathcal{B}_c^{\gamma\prime}$ by a matrix $g_j^{\gamma} \in \mathbb{G}^{\gamma}$, i.e. $g_i^{\gamma} = [\mathcal{B}_c^{\gamma} \to \mathcal{B}_c^{\gamma\prime}] \, g_j^{\gamma} [\mathcal{B}_c^{\gamma\prime} \to \mathcal{B}_c^{\gamma}] = \mathbf{F}_c^{\gamma} \, g_j^{\gamma} \, (\mathbf{F}_c^{\gamma})^{-1}$. These symmetries are thus globally preserved by the distortion; they belong to the subgroup:

$$\mathbb{H}_{\text{D}}^{\gamma} = \mathbb{G}^{\gamma} \cap \mathbf{F}_c^{\gamma} \, \mathbb{G}^{\gamma} \, (\mathbf{F}_c^{\gamma})^{-1} \qquad (30)$$

The distorted-shape variants are the cosets $d_i = g_i^{\gamma} \mathbb{H}_{\text{D}}^{\gamma}$, and the associated distortion matrices are $\mathbf{F}_c^{\gamma \to \alpha_i} = g_i^{\gamma} \, \mathbf{F}_c^{\gamma} (g_i^{\gamma})^{-1}$ with $g_i^{\gamma} \in d_i$. The number of variants is

$$N_{\text{D}}^{\alpha} = \frac{|\mathbb{G}^{\gamma}|}{|\mathbb{H}_{\text{D}}^{\gamma}|} \qquad (31)$$

If we consider the reverse transformation, the group of symmetries left invariant is now $\mathbb{H}_{\text{D}}^{\alpha} = \mathbb{G}^{\alpha} \cap \mathbf{F}_c^{\alpha} \, \mathbb{G}^{\alpha} \, (\mathbf{F}_c^{\alpha})^{-1}$ and the number of distorted-shape variants is $N_{\text{D}}^{\gamma} = \frac{|\mathbb{G}^{\alpha}|}{|\mathbb{H}_{\text{D}}^{\alpha}|}$.

In a previous paper (Cayron, 2016), we confused the "distorted-shape" variants with the "distortion" variants. However, equations (26) and (30) are not exactly the same; thus, apparently, the two types of variants should be distinguished. Mathematically, in the general case, $\mathbb{H}_{\text{F}}^{\gamma} \leq \mathbb{H}_{\text{D}}^{\gamma}$. Indeed, $g_i^{\gamma}$ belongs to $\mathbb{H}_{\text{D}}^{\gamma}$ if it exists a matrix $g_j^{\gamma} \in \mathbb{G}^{\gamma}$ such that $\mathbf{F}_c^{\gamma} \, g_i^{\gamma} \, (\mathbf{F}_c^{\gamma})^{-1} = g_j^{\gamma}$, and $g_i^{\gamma}$ belongs to $\mathbb{H}_{\text{F}}^{\gamma}$ for the same reason, but with the additional condition that $g_j^{\gamma} = g_i^{\gamma}$. This implies that the number of distorted-shape variants is lower than, or equal to, the number of distortion variants.

Physically, as the matrix $\mathbf{F}_c^{\gamma}$ is close to identity, it is difficult to imagine cases in which the symmetry matrix $g_i^{\gamma}$ would be changed into a different symmetry matrix $g_j^{\gamma}$. An example of such odd distortions will be given in § 7.2; the finish matrix $\mathbf{F}(\theta_f)$ that exchanges some symmetry elements has a negative determinant, which means that there exists a crossing a point (a certain value of angle $\theta$) for which $Det(\mathbf{F}(\theta)) = 0$. As the determinant gives the ratio of volume change during the transformation, the lattice should "disappear". Such continuous distortions are "physically" unrealistic and will not be considered anymore. We assume that for any structural displacive transformation, $\mathbb{H}_{\text{F}}^{\gamma} = \mathbb{H}_{\text{D}}^{\gamma}$, $N_{\text{F}}^{\alpha} = N_{\text{D}}^{\alpha}$, there is no distinction between the distortion variants and the distorted-shape variants.

### 5.2.3. Stretch variants

The stretch variants are calculated similarly as the distortion variants, but by replacing the distortion matrix $\mathbf{F}_c^\gamma$ in equation (26) by the stretch matrix $\mathbf{U}_c^\gamma$ defined by equation (2). The intersection group is

$$\mathbb{H}_U^\gamma = \{ g_i^\gamma \in \mathbb{G}^\gamma, \ g_i^\gamma \, \mathbf{U}_c^\gamma \left(g_i^\gamma\right)^{-1} = \mathbf{U}_c^\gamma\} \tag{32}$$

Similarly as in §5.2.1, $\mathbb{H}_U^\gamma \leq \mathbb{G}^\gamma$ is the stabilizer of $\mathbf{U}_c^\gamma$ in $\mathbb{G}^\gamma$. The number of variants is

$$N_U^\alpha = \frac{|\mathbb{G}^\gamma|}{|\mathbb{H}_U^\gamma|} \tag{33}$$

By considering equations (26) and (32), we show that $\mathbb{H}_F^\gamma \leq \mathbb{H}_U^\gamma$. The demonstration is easier by writing all the matrices in the orthonormal basis $\mathcal{B}_\#^\gamma$, i.e. by working with the isomorphic subgroups

$$\mathbb{H}_{F\#}^\gamma = \{ g_{i\#}^\gamma \in \mathbb{G}_\#^\gamma, \ g_{i\#}^\gamma \, \mathbf{F}_\#^\gamma \left(g_{i\#}^\gamma\right)^{-1} = \mathbf{F}_\#^\gamma\} = \mathcal{S}^\gamma \, \mathbb{H}_F^\gamma \, \mathcal{S}^{\gamma \, -1}, \text{ and } \mathbb{H}_{U\#}^\gamma = \mathcal{S}^\gamma \, \mathbb{H}_U^\gamma \, \mathcal{S}^{\gamma \, -1}.$$

If $g_{i\#}^\gamma \in \mathbb{H}_{F\#}^\gamma$, $g_{i\#}^\gamma \, \mathbf{F}_\#^\gamma \left(g_{i\#}^\gamma\right)^{-1} = \mathbf{F}_\#^\gamma$. By using $\left(g_{i\#}^\gamma\right)^{-1} = \left(g_{i\#}^\gamma\right)^t$, it comes that $g_{i\#}^\gamma \left(\mathbf{F}_\#^\gamma\right)^t \mathbf{F}_\#^\gamma \left(g_{i\#}^\gamma\right)^{-1} = \left(\mathbf{F}_\#^\gamma\right)^t \mathbf{F}_\#^\gamma$, thus $g_{i\#}^\gamma \left(\mathbf{U}_\#^\gamma\right)^2 \left(g_{i\#}^\gamma\right)^{-1} = \left(\mathbf{U}_\#^\gamma\right)^2$. As the matrix $\mathbf{U}_\#^\gamma$ is symmetric, there exists an orthonormal basis $\mathcal{B}_\Delta^\gamma$ in which $\mathbf{U}_\#^\gamma = \mathbf{R}^{-1} \, \mathbf{U}_\Delta^\gamma \, \mathbf{R}$ with $\mathbf{U}_\Delta^\gamma$ is diagonal and $\mathbf{R} = [\mathcal{B}_\Delta^\gamma \to \mathcal{B}_\#^\gamma]$ is a rotation. It comes that $g_{i\Delta}^\gamma \left(\mathbf{U}_\Delta^\gamma\right)^2 \left(g_{i\Delta}^\gamma\right)^{-1} = \left(\mathbf{U}_\Delta^\gamma\right)^2$, which imposes that $g_{i\Delta}^\gamma \, \mathbf{U}_\Delta^\gamma \left(g_{i\Delta}^\gamma\right)^{-1} = \mathbf{U}_\Delta^\gamma$ because $\mathbf{U}_\Delta^\gamma$ a diagonal matrix made of positive real numbers (in the general case $\mathbf{A} \neq \mathbf{B}$ does not necessarily imply that $\mathbf{A}^2 \neq \mathbf{B}^2$). Consequently, $g_\#^\gamma \, \mathbf{U}_\#^\gamma \left(g_\#^\gamma\right)^{-1} = \mathbf{U}_\#^\gamma$, i.e. $g_\#^\gamma \in \mathbb{H}_{U\#}^\gamma$. This proves that $\mathbb{H}_{F\#}^\gamma \leq \mathbb{H}_{U\#}^\gamma$, thus $\mathbb{H}_F^\gamma \leq \mathbb{H}_U^\gamma$, which implies that the numbers of distortion and stretch variants obey the inequality $N_U^\alpha \leq N_F^\alpha$. An example will be given in § 7.3.

### 5.3. Correspondence variants

The correspondence variants are mathematically defined with the subgroup $\mathbb{H}_C^\gamma$ of the parent symmetries that are in correspondence with daughter symmetries, even if the symmetry elements do not coincide. More precisely

$$\mathbb{H}_C^\gamma = \mathbf{C}_c^{\gamma \to \alpha} \mathbb{G}^\alpha \left(\mathbf{C}_c^{\gamma \to \alpha}\right)^{-1} \tag{34}$$

with $\mathbf{C}_c^{\gamma \to \alpha}$ the correspondence matrix, and $\mathbb{G}^\gamma$ and $\mathbb{G}^\alpha$ the point groups of parent and daughter phases. As for the orientation variants, the correspondence variants are defined by the cosets $c_i^\gamma = g_i^\gamma \mathbb{H}_C^\gamma$ and their correspondence matrices are $\mathbf{C}_c^{\gamma \to \alpha_i} = g_i^\gamma \mathbf{C}_c^{\gamma \to \alpha}$ with $g_i^\gamma \in c_i^\gamma$ (matrices in the coset $c_i^\gamma$). All the matrices that belong to $\mathbb{H}_C^\gamma$ point to the same correspondence variant $c_1^\gamma$, all the matrices that belong to $g_2^\gamma \mathbb{H}_C^\gamma$, with $g_2^\gamma \notin c_1^\gamma$, point to the same correspondence variant $c_2^\gamma$, etc. The number of correspondence variants is

$$N_C^\alpha = \frac{|\mathbb{G}^\gamma|}{|\mathbb{H}_C^\gamma|} \tag{35}$$

Now, if we consider the same formula for the reverse transformation, we get $N_C^\gamma = \frac{|\mathbb{G}^\alpha|}{|\mathbb{H}_C^\alpha|}$. As $\mathbb{H}_C^\alpha$ and $\mathbb{H}_C^\gamma$ are linked by the isomorphism $\mathbb{H}_C^\gamma = \mathbf{C}_c^{\gamma\to\alpha} \mathbb{H}_C^\alpha \left(\mathbf{C}_c^{\gamma\to\alpha}\right)^{-1}$, their orders are equal, $|\mathbb{H}_C^\alpha| = |\mathbb{H}_C^\gamma|$; which is expected because both sides of the equation give the number of symmetries in correspondence between the two phases. It then comes immediately that the number of variants of the direct transformation and those of the reverse transformation are linked by the relation:

$$N_C^\alpha |\mathbb{G}^\alpha| = N_C^\gamma |\mathbb{G}^\gamma| \tag{36}$$

When there is an *correspondence* group-subgroup relation between the daughter and parent phases $\mathbf{C}_c^{\gamma\to\alpha} \mathbb{G}^\alpha \left(\mathbf{C}_c^{\gamma\to\alpha}\right)^{-1} \subset \mathbb{G}^\gamma$, thus $\mathbb{H}_C^\gamma = \mathbf{C}_c^{\gamma\to\alpha} \mathbb{G}^\alpha \left(\mathbf{C}_c^{\gamma\to\alpha}\right)^{-1} \equiv \mathbb{G}^\alpha$, and $N_C^\alpha = \frac{|\mathbb{G}^\gamma|}{|\mathbb{G}^\alpha|}$. The number of parent orientations created by the reverse transformation of martensite variants is $N_C^\gamma = \frac{|\mathbb{G}^\alpha|}{|\mathbb{H}_T^\alpha|} = 1$, i.e. there is no new correspondence created by cycles of transformation.

The important difference between the *orientation* group-subgroup relation and the *correspondence* group-subgroup relation results from the nature of the symmetries that are considered, as it will be detailed in § 5.4.3.

## 5.4. The differences between the types of variants

### 5.4.1. No link systematic between the correspondence and stretch variants

For fcc-bcc transformation in steels, the OR between martensite and austenite that is usually observed is KS. The number of common symmetry elements is only 2 (Identity and Inversion), which implies by equation (24) that the number of orientation variants is $N_T = 48/2 = 24$. As Identity and Inversion are also the unique symmetries preserved by the lattice distortion, the number of distortion variants is also $N_F = N_D = 24$. The stretch distortion **U** follows Bain's model; it consists in a contraction along a $<001>_\gamma$ direction and dilatations along the two $<110>_\gamma$ directions normal to the contraction axis. The choice of the contraction axis determines the stretch variant; thus $N_U = 3$. The number of correspondence variants $N_C$ is given by equations (34) and (35) with $\mathbb{G}^\gamma$ and $\mathbb{G}^\alpha$ the m$\bar{3}$m cubic point group made of 48 symmetry matrices and $\mathbf{C}_c^{\gamma\to\alpha} = \begin{pmatrix} 1/2 & -1/2 & 0 \\ 1/2 & 1/2 & 0 \\ 0 & 0 & 1 \end{pmatrix}$; it is $N_C = 3$. In this case, $N_U = 3$ and $N_C = 3$,cbut it is important to keep in mind that the equality $N_C = N_U$ is fortuitous. We have already shown in § 2.3 that there is no general one-one relation between the correspondence matrix and the stretch matrix; so there is no systematic rule that would allow stating that $N_C = N_U$ in the general case. Actually counterexamples will be presented in § 7.4 ($N_U = 2$ and $N_C = 4$). One may think that at least $N_C \leq N_T$, but this is again not true, as it will be shown in § 7.7.2 ($N_C = 4$ and $N_T = 2$).

**5.4.2. No systematic link between the distortion and orientation variants**

In the discussion made by Cayron (2016) page 434, we showed in the case of a fcc-hcp transformation with Shoji-Nishiyama OR that there are 12 distortion variants and 4 orientation variants. We assumed that the inequality $N_F \geq N_T$ was general, but we did not provide the demonstration. Despite many efforts we could not succeed to show that $\mathbb{H}_F^\gamma \leq \mathbb{H}_T^\gamma$, mainly because $\mathbb{H}_F^\gamma$ is defined only from the point group $\mathbb{G}^\gamma$, whereas $\mathbb{H}_T^\gamma$ is defined from $\mathbb{G}^\gamma$ and $\mathbb{G}^\alpha$, without a priori no specific systematic relation between the two point groups. Actually, the inequality $N_F \geq N_T$ is not always true. A simple counterexample is given by the ferroelectric transition in PbTiO3 cubic m$\bar{3}$m → tetragonal p4mm with the edge/edge <100>$_\gamma$ // <100>$_\alpha$ OR. The relative displacement (shuffle) of the positively and negatively charged atoms in the unit cell, the lattice distortion and the polarization are correlated phenomena. As the daughter phase is not centrosymmetric, the same distortion, here a pure stretch, can generate (or be generated by) two ferroelectric domains at 180°. The number of distortion variants is 3 (along the *x*,*y* and *z* axes) whereas the number of orientation variants is 6. The inequality $N_F \leq N_T$ actually holds for simple (correspondence or orientation) group-subgroup relations (see § 5.4.4). Another example with $N_F < N_T$ will be given in § 7.5.

**5.4.3. No systematic link between the orientation and correspondence variants**

The orientation and correspondence variants are based on a formula implying the point groups of parent and daughter phases via the intersection groups given in equations (23) and (34), and they differ only by the use of $\mathbf{T}_c^{\gamma \to \alpha}$ and $\mathbf{C}_c^{\gamma \to \alpha}$, respectively. To get a better understanding of the correspondence variants, let us consider the case of a simple correspondence $\mathbf{C}_c^{\gamma \to \alpha} = \mathbf{I}$. Equation (34) becomes simply $\mathbb{H}_C^\gamma = \mathbb{G}^\gamma \cap \mathbb{G}^\alpha$, which means the correspondence variants are based on the subgroup of common symmetries. However, the orientation variants are also based the subgroup of common symmetries $\mathbb{H}_T^\gamma = \mathbb{G}^\gamma \cap \mathbf{T}_c^{\gamma \to \alpha} \mathbb{G}^\alpha (\mathbf{T}_c^{\gamma \to \alpha})^{-1}$. The distinction between the two subgroups comes from the subtile but fundamental difference between the *geometric* symmetries of $\mathbb{H}_T^\gamma$ and the *algebraic* symmetries of $\mathbb{H}_C^\gamma$. A symmetry of γ phase belongs to the *orientation* subgroup $\mathbb{H}_T^\gamma$ when its geometric element (mirror plane, rotation axis etc.) coincides with a similar symmetry element of the α phase; the matrices that represent these identical elements are equal when expressed in the same basis (here $\mathcal{B}_c^\gamma$) thanks to $\mathbf{T}_c^{\gamma \to \alpha}$. A symmetry of γ phase belongs to the *correspondence* subgroup $\mathbb{H}_C^\gamma$ when its matrix is equal to the matrix of a symmetry of the α phase, the two matrices being expressed in their own crystallographic bases. These symmetries are identical from an algebraic point of view, even if they do not necessarily represent the same geometric element. In other words, the geometric symmetries are the usual symmetries we are familiar with, i.e. inversion, reflection, rotations; they have an absolute nature and can be defined independently of the crystal, and the algebraic symmetries are more abstract; their significances are based on the permutations of the basis vectors and depend on the point group of the crystal on which they operate. An example of abstract algebraic symmetry is given by Morley's theorem, nicely demonstrated by Connes (2004), stating that in *any* triangle, the three points of intersection of the adjacent angle trisectors form an equilateral triangle. Algebraic symmetries are usually introduced in the representation theory of finite groups. The difference between the geometric and algebraic symmetries for phase transformations is illustrated with the example $\gamma \rightarrow \alpha$ that will be detailed in § 7.3 where $\gamma$ is a square crystal and $\alpha$ a hexagonal crystal. The horizontal reflection is encoded by the matrix $\boldsymbol{g}_3^{Sq} = \begin{pmatrix} 1 & 0 \\ 0 & -1 \end{pmatrix}$ in basis $\boldsymbol{\mathcal{B}}_c^{Sq}$, and by the matrix $\boldsymbol{g}_{11}^{Hx} = \begin{pmatrix} 1 & -1 \\ 0 & -1 \end{pmatrix}$ in the basis $\boldsymbol{\mathcal{B}}_c^{Hx}$, and with the help of $\mathbf{T}_c^{Sq \rightarrow Hx} = \begin{pmatrix} 1 & -1/2 \\ 0 & \sqrt{3}/2 \end{pmatrix}$, this matrix is written $\mathbf{T}_c^{Sq \rightarrow Hx} \boldsymbol{g}_{11}^{Hx} \left(\mathbf{T}_c^{Sq \rightarrow Hx}\right)^{-1} = \begin{pmatrix} 1 & 0 \\ 0 & -1 \end{pmatrix}$ in $\boldsymbol{\mathcal{B}}_c^{Sq}$. The matrices $\boldsymbol{g}_3^{Sq}$ and $\boldsymbol{g}_{11}^{Hx}$ represent the same geometric symmetry element, even if they are algebraic different. In contrast, as $\mathbf{C}_c^{Sq \rightarrow Hx} = \mathbf{I}$, $\boldsymbol{g}_3^{Sq}$ cannot be put in correspondence with any symmetry $\boldsymbol{g}_i^{Hx}$ of the hexagonal phase. The unique matrices in correspondence are $\mathbf{I}$, $-\mathbf{I}$, $\begin{pmatrix} 0 & 1 \\ 1 & 0 \end{pmatrix}$ and $-\begin{pmatrix} 0 & 1 \\ 1 & 0 \end{pmatrix}$, i.e. $\boldsymbol{g}_1^{Sq} = \boldsymbol{g}_1^{Hx}$, $\boldsymbol{g}_4^{Sq} = \boldsymbol{g}_4^{Hx}$, $\boldsymbol{g}_5^{Sq} = \boldsymbol{g}_{10}^{Hx}$ and $\boldsymbol{g}_8^{Sq} = \boldsymbol{g}_7^{Hx}$, and this is true whatever the OR. The matrices $\boldsymbol{g}_5^{Sq} = \boldsymbol{g}_{10}^{Hx}$ for example represent the same algebraic operation that interchanges the basis vectors $(\mathbf{a}^{Sq} \rightarrow \mathbf{b}^{Sq}, \mathbf{b}^{Sq} \rightarrow \mathbf{a}^{Sq})$ or $(\mathbf{a}^{Hx} \rightarrow \mathbf{b}^{Hx}, \mathbf{b}^{Hx} \rightarrow \mathbf{a}^{Hx})$, but obviously the geometric elements are different as the mirror plane is at 45° in the square phase, and 60° in the hexagonal phase. As the orientation and correspondence variants rely on different concepts, there are no systematic equalities or even inequalities between the number of orientation and correspondence variants. Some examples where $N_T \neq N_C$ will be shown in §7.

### 5.4.4. Specific cases with simple correspondence and group-subgroup relations

We have shown that there are no systematic relations between $N_F$, $N_C$ and $N_T$, the numbers of variants of distortion, correspondence and orientation. Inequalities can be established only for specific cases. We consider here transformations with a simple correspondence associated with a) a *correspondence* group-subgroup relation, or b) an *orientation* group-subgroup relation.

a) For a *correspondence* group-subgroup relation (common *algebraic* symmetries), $\mathbb{G}^\alpha \leq \mathbb{G}^\gamma$. As $\mathbf{C}_c^{\gamma \rightarrow \alpha} = \mathbf{I}$, $\mathbb{H}_C^\gamma = \mathbb{G}^\alpha$, and $N_C^\alpha = \frac{|\mathbb{G}^\gamma|}{|\mathbb{G}^\alpha|}$. In addition, $\left| \mathbb{G}^\gamma \cap \mathbf{T}_c^{\gamma \rightarrow \alpha} \mathbb{G}^\alpha \left(\mathbf{T}_c^{\gamma \rightarrow \alpha}\right)^{-1} \right| \leq |\mathbb{G}^\alpha|$, which implies that $|\mathbb{H}_T^\gamma| \leq |\mathbb{H}_C^\gamma|$, and thus $N_C^\alpha \leq N_T^\alpha$. In addition, $\mathbf{F}_c^\gamma = \mathbf{T}_c^{\gamma \rightarrow \alpha}$, $\mathbb{G}^\gamma \cap \mathbf{T}_c^{\gamma \rightarrow \alpha} \mathbb{G}^\alpha \left(\mathbf{T}_c^{\gamma \rightarrow \alpha}\right)^{-1} \leq \mathbb{G}^\gamma \cap \mathbf{F}_c^\gamma \mathbb{G}^\gamma \left(\mathbf{F}_c^\gamma\right)^{-1}$, i.e. $\mathbb{H}_T^\gamma \leq \mathbb{H}_F^\gamma$, and thus $N_F^\alpha \leq N_T^\alpha$. Consequently the inequalities are $N_C^\alpha \leq N_T^\alpha$ and $N_F^\alpha \leq N_T^\alpha$, without systematic inequality between $N_C^\alpha$ and $N_F^\alpha$.

b) For an *orientation* group-subgroup relation (common *geometric* symmetries), $\mathbf{T}_c^{\gamma \rightarrow \alpha} \mathbb{G}^\alpha \left(\mathbf{T}_c^{\gamma \rightarrow \alpha}\right)^{-1} \leq \mathbb{G}^\gamma$. Thus, $\mathbb{H}_T^\gamma = \mathbf{T}_c^{\gamma \rightarrow \alpha} \mathbb{G}^\alpha \left(\mathbf{T}_c^{\gamma \rightarrow \alpha}\right)^{-1} \equiv \mathbb{G}^\alpha$, and $N_T^\alpha = \frac{|\mathbb{G}^\gamma|}{|\mathbb{G}^\alpha|}$. As $\mathbf{C}_c^{\gamma \rightarrow \alpha} = \mathbf{I}$, $\mathbb{H}_C^\gamma = \mathbb{G}^\gamma \cap \mathbb{G}^\alpha$, thus $|\mathbb{H}_C^\gamma| \leq |\mathbb{G}^\alpha| = |\mathbb{H}_T^\gamma|$, and $N_T^\alpha \leq N_C^\alpha$. In addition, as $\mathbf{F}_c^\gamma = \mathbf{T}_c^{\gamma \rightarrow \alpha}$, $\mathbb{G}^\gamma \cap$

$\mathbf{T}_c^{\gamma\to\alpha}\mathbb{G}^\alpha\left(\mathbf{T}_c^{\gamma\to\alpha}\right)^{-1} \leq \mathbb{G}^\gamma \cap \mathbf{F}_c^\gamma \mathbb{G}^\gamma\left(\mathbf{F}_c^\gamma\right)^{-1}$, i.e. $\mathbb{H}_T^\gamma \leq \mathbb{H}_F^\gamma$, and thus $N_F^\alpha \leq N_T^\alpha$. Consequently, $N_F^\alpha \leq N_T^\alpha \leq N_C^\alpha$. A typical example is the PbTiO$_3$ transition cubic m$\bar{3}$m $\to$ tetragonal p4mm with a parallelism of the directions <100> mentioned in § 5.4.2 ($N_F^\alpha = 3$, $N_T^\alpha = N_C^\alpha = 6$).

## 6. Orientation variants by cycles of transformations

As mentioned in § 1.3, the reasons of reversibility/irreversibility during thermal cycling are not yet fully understood. A part of irreversibility comes from the accumulation of defects (dislocations) generated by series of transformation, which was mathematically described with the "global" group (Bhattacharya *et al.*, 2004) and Cayley graphs (Gao *et al.*, 2017). Variant pairing/grouping whose details depend on compatibility conditions allows reducing the amount of dislocations (James & Hane, 2000; Bhattacharya *et al.*, 2004; Gao *et al.* 2016, 2017). Another part of irreversibility is intrinsically due to the orientation symmetries, independently of the lattice parameters and metrics of the parent and daughter phases. We have seen in § 5.1 that when an orientation group-subgroup relationship exists between the parent and daughter crystals, then $N_T^\gamma = \frac{|\mathbb{G}^\alpha|}{|\mathbb{H}_T^\alpha|} = 1$, which means that there is only one possible orientation of γ of second generation (after one cycle) $\gamma^1 \to \{\alpha^1\} \to \{\gamma^2\}$. The cycling graph is finite. Inversely, in absence of orientation group-subgroup relation, new orientations are created, at least after one cycle. By generalizing the orientation variants defined in § 5.1, it can be show that after *n* cycles $\gamma^1 \to \{\alpha^1\} \to \{\gamma^2\} \to \ldots \to \{\alpha^n\} \to \{\gamma^n\}$ the γ variants at the $n^{\text{th}}$ generation can be defined by a chain of type $g_i^\gamma \mathbb{H}_T^\gamma \mathbf{T}^{\gamma\to\alpha} \cdot g_k^\alpha \mathbb{H}_T^\alpha \mathbf{T}^{\alpha\to\gamma} \ldots g_m^\gamma \mathbb{H}_T^\gamma \mathbf{T}^{\gamma\to\alpha} \cdot g_n^\alpha \mathbb{H}_T^\alpha \mathbf{T}^{\alpha\to\gamma}$, with $\mathbf{T}^{\alpha\to\gamma} = \left(\mathbf{T}^{\gamma\to\alpha}\right)^{-1}$ and $\mathbb{H}_T^\alpha \mathbf{T}^{\alpha\to\gamma} = \mathbf{T}^{\alpha\to\gamma}\mathbb{H}_T^\gamma$. The *n*-chains are thus *n*-cosets of type $g_i^\gamma \mathbb{H}_T^\gamma g_{k/\gamma}^\alpha \mathbb{H}_T^\gamma \ldots g_m^\gamma \mathbb{H}_T^\gamma g_{n/\gamma}^\alpha \mathbb{H}_T^\gamma$, where $g_{k/\gamma}^\alpha = \mathbf{T}^{\gamma\to\alpha} g_k^\alpha \left(\mathbf{T}^{\gamma\to\alpha}\right)^{-1}$ are symmetry operations of the α phase written in the γ basis. The number of simple cosets is given by Lagrange's formula, the number of double-cosets by Burnside's formulae (Cayron, 2006), but to our knowledge, there is not yet a general formula that gives the number of *n*-cosets; the mathematical problem seems open. In the specific case of $\Sigma 3^n$ multiple twinning, the structure of the *n*-variants was geometrically represented by a fractal Cayley graph, and algebraically modelled by strings associated with a concatenation rule that effectively replaces matrix multiplication; the algebraic structure depends on the choice of the representatives in the simple cosets forming the *n*-cosets: it can be a free group (Reed *et al*, 2004) or a semi-group (Cayron, 2007); actually, the general structure is a groupoid (Cayron, 2006, 2007). The difficulty in finding in the general case a formula for N(*n*), the number of γ orientation variants after *n*-cycles, is probably partially due to the "flexibility" of this mathematical structure. Obviously, N$(n) \leq N_T^\alpha N_T^\gamma N_T^\alpha N_T^\gamma \ldots n\ times$. In some special cases this number can become constant after a limited number of cycles, the graph is finite. It is the case when there is an orientation group-subgroup relation; the maximum number of variants is reached at *n* = 1. One can imagine cases where the saturation is reached at higher *n*; a 2D example of saturation at *n* = 2 will be shown in § 7.9. In the case of $\sum 3^n$ multiple twinning, N($n$) = 4.3$^n$, the graph of orientation variants is infinite. Another simple 2D example of infinite graph will be shown in § 7.10. This 2D example will also prove that the graph of orientation variants related to fcc-bcc transformation with a KS OR is infinite, which probably partially explain why fcc-bcc martensitic transformation in steels is irreversible.

A complete crystallographic theory of transformation cycling taking into account all the aspects of the transformation (compatibility, variant grouping, accumulated defect, orientation reversibility) in a unified mathematical framework is still the subject of intense research in different labs all over the world.

## 7. Examples of variants with 2D transformations

This section gives 2D examples that should help the reader to understand the notions of orientation, distortion and correspondence matrices, and their variants. They imply square, rectangular, hexagonal, and triangular "phases" whose symmetries form the point groups noted $\mathbb{G}^{Sq}$, $\mathbb{G}^{Rc}$, $\mathbb{G}^{Hx}$, $\mathbb{G}^{Tr}$, respectively. These groups are explicitly given by the sets of 2x2 matrices reported in Appendix C. The distortion and orientation variants are graphically represented, but not the correspondence variants because of their algebraic and non-geometric nature. The "real" 3D cases of displacive transformations and deformation twinning with fcc, bcc and hcp phases are technically more complex, but relies on the same notions; some are treated in the special case of hard-sphere hypothesis (Cayron, 2016, 2017a, 2017b, 2018).

### 7.1. Square *p4mm* → Square *p4mm* in Σ1 OR by simple strain

Let us consider the oversimplified example of a transformation from a 2D square crystal (*p4mm*) to the same square crystal (*p4mm*) generated by a simple strain, as illustrated in Figure 4. Here, $\mathbb{G}^\gamma = \mathbb{G}^\alpha = \mathbb{G}^{Sq}$, the group of eight matrices reported in Appendix C. The distortion matrix is $\mathbf{F}_c^\gamma = \begin{pmatrix} 1 & -1 \\ 0 & 1 \end{pmatrix}$, the orientation matrix is $\mathbf{T}_c^{\gamma \to \alpha} = \mathbf{I}$, and the correspondence matrix is $\mathbf{C}_c^{\alpha \to \gamma} = \begin{pmatrix} 1 & -1 \\ 0 & 1 \end{pmatrix}$. The distortion matrix $\mathbf{F}_c^\gamma$ is a lattice-preserving strain; it belongs to the "global group" (Battacharya *et al*., 2004; Gao *et al*. 2016, 2017). The distortion leaves globally invariant the lattice, and all the geometric symmetry elements are put in coincidence. However, some abstract symmetries are lost by the distortion and by the correspondence. The calculations show that $\mathbb{H}_F^\gamma = \mathbb{H}_D^\gamma = \mathbb{H}_C^\gamma = \{g_1^{Sq}, g_4^{Sq}\}$, thus $N_F^\alpha = N_D^\alpha = N_C^\alpha = 4$, and that $\mathbb{H}_T^\gamma = \mathbb{G}^\gamma$, thus $N_T^\alpha = 1$. By left polar decomposition $\mathbf{F}_c^\gamma = \mathbf{F}_\#^\gamma = \mathbf{Q}_\#^\gamma \mathbf{U}_\#^\gamma$ with $\mathbf{Q}_\#^\gamma = \begin{pmatrix} \frac{2}{\sqrt{5}} & \frac{-1}{\sqrt{5}} \\ \frac{1}{\sqrt{5}} & \frac{2}{\sqrt{5}} \end{pmatrix}$ and $\mathbf{U}_\#^\gamma = \begin{pmatrix} \frac{2}{\sqrt{5}} & \frac{-1}{\sqrt{5}} \\ \frac{-1}{\sqrt{5}} & \frac{3}{\sqrt{5}} \end{pmatrix}$ whose eigenvalues are $\left\{\frac{5+\sqrt{5}}{2\sqrt{5}}, \frac{5-\sqrt{5}}{2\sqrt{5}}\right\} \approx$ {1.62,0.62} along the eigenvectors $\left[\frac{1}{2}(1-\sqrt{5}), 1\right]^t$ and $\left[\frac{1}{2}(1+\sqrt{5}), 1\right]^t$, i.e. an expansion / contraction along these two vectors. The calculations also show that $\mathbb{H}_U^\gamma = \{g_1^{Sq}, g_4^{Sq}\}$, thus $N_U^\alpha = 4$. The distinct stretch matrices are $\begin{pmatrix} \frac{2}{\sqrt{5}} & \frac{1}{\sqrt{5}} \\ \frac{1}{\sqrt{5}} & \frac{3}{\sqrt{5}} \end{pmatrix}$, $\begin{pmatrix} \frac{3}{\sqrt{5}} & \frac{1}{\sqrt{5}} \\ \frac{1}{\sqrt{5}} & \frac{2}{\sqrt{5}} \end{pmatrix}$, $\begin{pmatrix} \frac{2}{\sqrt{5}} & \frac{-1}{\sqrt{5}} \\ \frac{-1}{\sqrt{5}} & \frac{3}{\sqrt{5}} \end{pmatrix}$, and $\begin{pmatrix} \frac{3}{\sqrt{5}} & \frac{-1}{\sqrt{5}} \\ \frac{-1}{\sqrt{5}} & \frac{2}{\sqrt{5}} \end{pmatrix}$. Different continuous forms of the distortion can be proposed, such as $\mathbf{F}_c^\gamma(\beta) = \begin{pmatrix} 1 & \text{Tan}(\frac{\pi}{2} - \beta) \\ 0 & 1 \end{pmatrix}$ with $\beta \in [\frac{\pi}{2} \to \frac{\pi}{4}]$, which corresponds to a continuous simple strain. Please note that the atoms do not remain in contact during this type of distortion; the intermediate states are thus probably energetically not realistic. This example was used only as a mathematical example of transformation with $N_T^\alpha = 1$. It would be better treated with discrete discontinuous dislocation gliding than with continuous lattice distortion.

### 7.2. Square *p4mm* → Square *p4mm* in Σ1 OR by turning inside out

This example, shown in Figure 5, is also purely mathematic. It uses the same parent and daughter square crystal as in the previous example, i.e. $\mathbb{G}^\gamma = \mathbb{G}^\alpha = \mathbb{G}^{Sq}$, with the same square-square OR, but now it imagines that the distortion matrix is $\mathbf{F}_c^\gamma = \begin{pmatrix} 0 & -1 \\ -1 & 0 \end{pmatrix}$. This distortion is very special because it reverses the handedness of the basis. The matrix $\mathbf{F}_c^\gamma$ can be diagonalized in a basis rotated by $\frac{\pi}{4}$, it is $\mathbf{F}_{\pi/4}^\gamma = \begin{pmatrix} 1 & 0 \\ 0 & -1 \end{pmatrix}$. The orientation matrix is $\mathbf{T}_c^{\gamma \to \alpha} = \mathbf{I}$, and the correspondence matrix is $\mathbf{C}_c^{\alpha \to \gamma} = \begin{pmatrix} 0 & -1 \\ -1 & 0 \end{pmatrix}$. The calculations show that $\mathbb{H}_F^\gamma = \{g_1^{Sq}, g_4^{Sq}, g_5^{Sq}, g_8^{Sq}\}$, thus $N_F^\alpha = 2$, and that $\mathbb{H}_D^\gamma = \mathbb{H}_C^\gamma = \mathbb{H}_T^\gamma = \mathbb{G}^\gamma$, thus $N_D^\alpha = N_C^\alpha = N_T^\alpha = 1$. This is the only example of the section in which $N_F^\alpha \neq N_D^\alpha$. This case is not "physical" because it implies turning inside out the surface of a circle, which is impossible in 2D (notice that it is possible in 3D, it is the famous sphere eversion "paradox"). To illustrate this point, let us consider a possible continuous form of this distortion $\mathbf{F}_c^\gamma(\beta) = \begin{pmatrix} \text{Cos}(\frac{\beta-\pi/2}{2}) & -\text{Sin}(\frac{\beta-\pi/2}{2}) \\ -\text{Sin}(\frac{\beta-\pi/2}{2}) & \text{Cos}(\frac{\beta-\pi/2}{2}) \end{pmatrix}$ with $\beta \in [\frac{\pi}{2} \to \frac{3\pi}{2}]$. At $\beta = \pi$, the determinant of the distortion is zero, i.e. the lattice becomes flat, as shown in Figure 5b. More generally, as the determinant is a multilinear function of the matrix coefficients, any continuous path from the starting state, $\mathbf{F}_c^\gamma(\beta_s) = \mathbf{I}$, to the final state, $\mathbf{F}_c^\gamma(\beta_f) = \mathbf{F}_c^\gamma$, with $\text{Det}(\mathbf{F}_c^\gamma) = -1$, implies that it exists an intermediate at the angle $\beta_i$ such that $\text{Det}(\mathbf{F}_c^\gamma(\beta_i)) = 0$; which can be considered as "non-crystallographic" because the lattice loses one of its dimension.

### 7.3. Square *p4mm* → Hexagon *p6mm* by angular distortion

In this example, the square parent crystal is transformed into a hexagonal daughter crystal, as shown in Figure 6. The parent and daughter symmetries are given by $\mathbb{G}^\gamma = \mathbb{G}^{Sq}$ and $\mathbb{G}^\alpha = \mathbb{G}^{Hx}$, respectively. Once the distortion is complete the distortion matrix is $\mathbf{F}_c^\gamma = \begin{pmatrix} 1 & \frac{-1}{2} \\ 0 & \frac{\sqrt{3}}{2} \end{pmatrix}$. The orientation matrix is also $\mathbf{T}_c^{\gamma\to\alpha} = \begin{pmatrix} 1 & \frac{-1}{2} \\ 0 & \frac{\sqrt{3}}{2} \end{pmatrix}$, and the correspondence matrix is $\mathbf{C}_c^{\alpha\to\gamma} = \mathbf{I}$. The calculations show that $\mathbb{H}_F^\gamma = \mathbb{H}_D^\gamma = \{g_1^{Sq}, g_4^{Sq}\}$, $\mathbb{H}_T^\gamma = \{g_1^{Sq}, g_4^{Sq}, g_2^{Sq}, g_3^{Sq}\}$, $\mathbb{H}_C^\gamma = \{g_1^{Sq}, g_4^{Sq}, g_5^{Sq}, g_8^{Sq}\}$, and thus $N_F^\alpha = N_D^\alpha = 4$, and $N_T^\alpha = N_C^\alpha = 2$. Note that the intersection group defining the orientation variants is different from that of the correspondence variants. The left polar decomposition gives $\mathbf{F}_c^\gamma = \mathbf{F}_\#^\gamma = \mathbf{Q}_\#^\gamma \mathbf{U}_\#^\gamma$ with $\mathbf{U}_\#^\gamma = \frac{1}{4}\begin{pmatrix} \sqrt{2}+\sqrt{6} & \sqrt{2}-\sqrt{6} \\ \sqrt{2}-\sqrt{6} & \sqrt{2}+\sqrt{6} \end{pmatrix}$ and $\mathbf{Q}_\#^\gamma = \frac{1}{4}\begin{pmatrix} \sqrt{2}+\sqrt{6} & \sqrt{2}-\sqrt{6} \\ \sqrt{6}-\sqrt{2} & \sqrt{2}+\sqrt{6} \end{pmatrix}$ whose eigenvalues are $\left\{\sqrt{\frac{3}{2}}, \frac{1}{\sqrt{2}}\right\} \approx \{1.22, 0.71\}$ along the eigenvectors $[-1,1]^t$ and $[1,1]^t$, which geometrically means that the square is transformed into a rhombus by extension/contraction along its two diagonals. The calculations show that indeed $\mathbb{H}_U^\gamma = \{g_1^{Sq}, g_4^{Sq}, g_5^{Sq}, g_8^{Sq}\}$, and thus $N_U^\alpha = 2$; the two stretch matrices are $\frac{1}{4}\begin{pmatrix} \sqrt{2}+\sqrt{6} & \sqrt{2}-\sqrt{6} \\ \sqrt{2}-\sqrt{6} & \sqrt{2}+\sqrt{6} \end{pmatrix}$ and $\frac{1}{4}\begin{pmatrix} \sqrt{2}+\sqrt{6} & \sqrt{6}-\sqrt{2} \\ \sqrt{6}-\sqrt{2} & \sqrt{2}+\sqrt{6} \end{pmatrix}$. By assuming that the crystal is made of hard-disk, the distortion can be expressed by a continuous form $\mathbf{F}_c^\gamma(\beta) = \begin{pmatrix} 1 & \cos(\beta) \\ 0 & \sin(\beta) \end{pmatrix}$ with $\beta \in [\frac{\pi}{2} \to \frac{2\pi}{3}]$. This is a typical example of angular distortion. The inverse transformation will be considered in § 7.6.

### 7.4. Square *p4mm* → Triangle *p3m* by angular distortion

In this example, the distortion, orientation and correspondence matrices are exactly the same as in the previous example, but the square parent crystal now contains extra interstitial atoms, as shown in Figure 7. After distortion, these extra atoms cannot stay in the centre of the distorted cells because of steric reasons, and they move in one of the two hexagonal nets that are formed by the distortion; they are said to "shuffle". If the choice between the two possible sites is random, the final phase is hexagonal p6mm as in the previous example, but if one of the two sites is selected, by a domino effect for example, then the daughter phase is p3m1 and its symmetries are reduced to $\mathbb{G}^\alpha = \mathbb{G}^{Tr}$, as shown in this example. Since the distortion and the parent point group are the same as in the previous example, $\mathbb{H}_F^\gamma = \mathbb{H}_D^\gamma = \{g_1^{Sq}, g_4^{Sq}\}$ and $\mathbb{H}_U^\gamma = \{g_1^{Sq}, g_4^{Sq}, g_5^{Sq}, g_8^{Sq}\}$, thus $N_F^\alpha = N_D^\alpha = 4$ and $N_U^\alpha = 2$. The calculations with the daughter point group $\mathbb{G}^{Tr}$ show however a difference with the previous case: $\mathbb{H}_T^\gamma = \mathbb{H}_C^\gamma = \{g_1^{Sq}, g_4^{Sq}\}$, and thus $N_T^\alpha = N_C^\alpha = 4$. The four orientation variants are similar to those we used to illustrate the groupoid structure of orientation variants by Cayron (2006). This example is a case where $N_C^\alpha > N_U^\alpha$.

### 7.5. Square *p4mm* → Triangle *p3m* by pure stretch distortion

The parent and daughter phases are the same as in the previous example. The only difference comes from the distortion; it is now a pure stretch constituted of a contraction along one of the diagonal of the square and an elongation along the second diagonal, as illustrated in Figure 8. Once the distortion is complete the distortion matrix, already calculated in the example 7.3, is

$\mathbf{F}_c^\gamma = \mathbf{U}_\#^\gamma = \frac{1}{4}\begin{pmatrix} \sqrt{2}+\sqrt{6} & \sqrt{2}-\sqrt{6} \\ \sqrt{2}-\sqrt{6} & \sqrt{2}+\sqrt{6} \end{pmatrix}$. Since the correspondence matrix is $\mathbf{C}_c^{\alpha \to \gamma} = \mathbf{I}$, the orientation matrix is $\mathbf{T}_c^{\gamma \to \alpha} = \mathbf{F}_c^\gamma = \mathbf{U}_\#^\gamma$. In this example, $\mathbb{H}_F^\gamma = \mathbb{H}_D^\gamma = \mathbb{H}_C^\gamma = \{g_1^{Sq}, g_4^{Sq}, g_5^{Sq}, g_8^{Sq}\}$ and $\mathbb{H}_T^\gamma = \{g_1^{Sq}, g_4^{Sq}\}$, thus $N_F^\alpha = N_D^\alpha = N_C^\alpha = 2$ and $N_T^\alpha = 4$. This example is a case where $N_T^\alpha > N_F^\gamma$.

### 7.6. Hexagon *p6mm* → Square *p4mm* by angular distortion

This transformation, shown in Figure 9, is the reverse of the transformation described in §7.3. The calculation of the continuous form of the distortion matrix was fully treated in section 6 of Cayron (2018). Here, to keep coherency with the names given in §7.6, we write α the hexagonal parent phase, and γ the square daughter phase. The coordinate transformation matrix is the inverse of $\mathbf{T}_c^{\gamma \to \alpha} = \begin{pmatrix} 1 & \frac{-1}{2} \\ 0 & \frac{\sqrt{3}}{2} \end{pmatrix}$; it is $\mathbf{T}_c^{\alpha \to \gamma} = \begin{pmatrix} 1 & \frac{1}{\sqrt{3}} \\ 0 & \frac{2}{\sqrt{3}} \end{pmatrix}$. The correspondence matrix remains $\mathbf{C}_c^{\alpha \to \gamma} = \mathbf{I}$. Equation (16) gives $\mathbf{F}_c^\alpha(\beta) = \mathbf{T}_c^{\alpha \to \gamma}\mathbf{F}_c^\gamma(\beta)$ with $\mathbf{F}_c^\gamma(\beta) = \begin{pmatrix} 1 & Cos(\beta) \\ 0 & Sin(\beta) \end{pmatrix}$, thus $\mathbf{F}_c^\alpha(\beta) = \begin{pmatrix} 1 & Cos(\beta)+\frac{Sin(\beta)}{\sqrt{3}} \\ 0 & \frac{2Sin(\beta)}{\sqrt{3}} \end{pmatrix}$, with $\beta \in [\frac{2\pi}{3} \to \frac{\pi}{2}]$. When written in $\mathcal{B}_\#^\alpha$ thanks to the structure tensor $\mathcal{S}^\alpha = [\mathcal{B}_\#^\alpha \to \mathcal{B}_c^\alpha] = \begin{pmatrix} 1 & \frac{-1}{2} \\ 0 & \frac{\sqrt{3}}{2} \end{pmatrix}$, the distortion matrix becomes $\mathbf{F}_\#^\alpha(\beta) = \begin{pmatrix} 1 & \frac{1+2Cos(\beta)}{\sqrt{3}} \\ 0 & \frac{2Sin(\beta)}{\sqrt{3}} \end{pmatrix}$ in agreement with (Cayron, 2018). One can notice that $\mathbf{F}_c^\alpha(\beta)$ is not the inverse of $\mathbf{F}_c^\gamma(\beta)$ because the matrices are not expressed in the same basis.

The matrix of complete transformation is $\mathbf{F}_c^\alpha = \mathbf{F}_c^\alpha(\frac{\pi}{2}) = \begin{pmatrix} 1 & \frac{1}{\sqrt{3}} \\ 0 & \frac{2}{\sqrt{3}} \end{pmatrix}$. The variants are calculated with parent and daughter symmetries $\mathbb{G}^\alpha = \mathbb{G}^{Hx}$ and $\mathbb{G}^\gamma = \mathbb{G}^{Sq}$, respectively; the distortion matrix is $\mathbf{F}_C^\alpha$, the orientation matrix is $\mathbf{T}_c^{\alpha \to \gamma}$, and the correspondence matrix is $\mathbf{C}_c^{\alpha \to \gamma} = \mathbf{I}$. The calculations show that $\mathbb{H}_F^\alpha = \mathbb{H}_D^\alpha = \{g_1^{Hx}, g_4^{Hx}\}$, $\mathbb{H}_T^\gamma = \{g_1^{Hx}, g_2^{Hx}, g_4^{Hx}, g_5^{Hx}\}$, $\mathbb{H}_C^\alpha = \{g_1^{Hx}, g_4^{Hx}, g_7^{Hx}, g_{10}^{Hx}\}$, and thus the number of variants of the transformation α→γ is $N_F^\gamma = N_D^\gamma = 6$, and $N_T^\gamma = N_C^\gamma = 3$. By considering the number of variants of the transformation γ→α found in § 7.3, i.e. $N_F^\alpha = N_D^\alpha = 4$, and $N_T^\alpha = N_C^\alpha = $ 2, one can check the validity of equations (25) (29) and (35), i.e. $N_T^\alpha |\mathbb{G}^\alpha| = N_T^\gamma |\mathbb{G}^\gamma| = 24$, $N_F^\alpha |\mathbb{G}^\alpha| = N_F^\gamma |\mathbb{G}^\gamma| = 48$, and $N_C^\alpha |\mathbb{G}^\alpha| = N_C^\gamma |\mathbb{G}^\gamma| = 24$.

### 7.7. Square p4mm → Square p4mm in Σ5 OR

#### 7.7.1. By pure shuffle

We consider here two square crystals of same phase with a symmetry group $\mathbb{G}^\gamma = \mathbb{G}^\alpha = \mathbb{G}^{Sq}$, and related to each other by the specific misorientation Σ5, as illustrated in Figure 10a. The number Σ is the volume (here area) of the coincidence site lattice (CSL) divided by the volume of one of the individual lattice (Grimmer, Bollmann & Warrington, 1974; Gratias & Portier, 1982). The misorientation gives no information about the distortion, and additional assumptions are required to establish a crystallographic model. In this example, it is supposed that the Σ5 CSL (in light green in Figure 10a) is undistorted, and that only the atoms in the supercell move (curled black arrows). This type of transformation is called "pure shuffle". In this example, $\mathbf{B}_{super}^{\gamma'} = \mathbf{B}_{super}^{\gamma} = \begin{pmatrix} 2 & -1 \\ 1 & 2 \end{pmatrix}$, and $\mathbf{B}_{super}^{\alpha} = \begin{pmatrix} 2 & 1 \\ -1 & 2 \end{pmatrix}$. Equations (7), (8) and (9) directly lead to $\mathbf{T}_c^{\gamma \to \alpha} = \mathbf{C}_c^{\gamma \to \alpha} = \frac{1}{5}\begin{pmatrix} 3 & -4 \\ 4 & 3 \end{pmatrix}$, and $\mathbf{F}_c^\gamma = \mathbf{I}$. The calculations show that $\mathbb{H}_F^\gamma = \mathbb{H}_D^\gamma = \mathbb{G}^\gamma$ and $\mathbb{H}_T^\gamma = \mathbb{H}_C^\gamma = \{g_1^{Sq}, g_4^{Sq}, g_6^{Sq}, g_7^{Sq}\}$, thus $N_F^\alpha = N_D^\alpha = 1$, and $N_T^\alpha = N_C^\alpha = 2$.

#### 7.7.2. By simple shear

A simple shear can lead to the same configuration (same phases and same misorientation), as illustrated in Figure 10b. Again, the OR between the two crystals is $\mathbf{T}_c^{\gamma \to \alpha} = \frac{1}{5}\begin{pmatrix} 3 & -4 \\ 4 & 3 \end{pmatrix}$; however now the distortion implies a different supercell given by $\mathbf{B}_{super}^\gamma = \begin{pmatrix} 2 & -1 \\ 1 & 0 \end{pmatrix}$, $\mathbf{B}_{super}^\alpha = \begin{pmatrix} 2 & -1 \\ -1 & 1 \end{pmatrix}$, and $\mathbf{B}_{super}^{\gamma'} \ne \mathbf{B}_{super}^{\gamma}$. Equation (8) actually leads to $\mathbf{B}_{super}^{\gamma'} = \mathbf{T}_c^{\gamma \to \alpha} \mathbf{B}_{super}^{\alpha}$, thus equation (7) gives $\mathbf{F}_c^\gamma = \mathbf{T}_c^{\gamma \to \alpha} \mathbf{B}_{super}^\alpha (\mathbf{B}_{super}^\gamma)^{-1} = \frac{1}{5}\begin{pmatrix} 7 & -4 \\ 1 & 3 \end{pmatrix}$ and $\mathbf{C}_c^{\alpha \to \gamma} = (\mathbf{T}_c^{\gamma \to \alpha})^{-1} \mathbf{F}_c^\gamma = \begin{pmatrix} 1 & 0 \\ -1 & 1 \end{pmatrix}$. The unique eigenvalue of $\mathbf{F}_c^\gamma$ is 1 and is associated with the eigenvector $\mathbf{u} = [2,1]_\gamma^t$. Actually one can check that $\mathbf{F}_c^\gamma$ is a simple strain along the direction $\mathbf{u}$, and its amplitude is $ArcTan((\mathbf{F}_c^\gamma - \mathbf{I})\frac{1}{\sqrt{5}}[1,\bar{2}]_\gamma^t = ArcTan\left(\frac{1}{\sqrt{5}}\right) \approx 24.1°$. The calculations show that $\mathbb{H}_F^\gamma = \mathbb{H}_D^\gamma = \mathbb{H}_C^\gamma = \{g_1^{Sq}, g_4^{Sq}\}$ and $\mathbb{H}_T^\gamma = \{g_1^{Sq}, g_4^{Sq}, g_6^{Sq}, g_7^{Sq}\}$, thus $N_F^\alpha = N_D^\alpha = N_C^\alpha = 4$ and $N_T^\alpha = 2$. A continuous form of the distortion could be determined by assuming that the atoms are hard-circles, and one would find that the analytical equation of the atomic trajectories are not those of a continuous simple strain; simple strain is obtained only when the transformation is complete; it is the strain between the initial and final states without considering the exact path in between.

### 7.8. Square *p4mm* → Rectangle *pm* by short-range order angular distortion

In this example, we consider a crystal made of two types of atoms A and B with different diameters, noted $d_A$ and $d_B$. At high temperature, the thermal activation weakens the interatomic bonds so much that the system is fully disordered. Equivalently said, the entropy part dominates the enthalpy part in the free energy such that the system gets a configuration of high symmetry in order to increase at maximum the number of microstates. The high temperature phase is a square crystal γ made of "mean" atoms presented in blue in Figure 11a. The lattice parameter is $a^\gamma = \frac{d_A + d_B}{2}$ and the symmetries are given by $\mathbb{G}^\gamma = \mathbb{G}^{Sq}$. At low temperature, one can imagine the case in which the atoms A tend to attract each other and the atoms B tend to repulse each other, such that the atoms A and B organise themselves to form the rectangular crystal. The ordered phase before and after distortion is show in Figure 11b and Figure 11c, respectively. It is highly probable that ordering and distortion occur simultaneously as they are both component of the change of atomic bonds with temperature. The daughter symmetries are given by $\mathbb{G}^\alpha = \mathbb{G}^{Rc}$. Here, the diameter of the atoms is supposed to be constant during the transformation. The new lattice is such that $a^\alpha = d_A$ and $b^\alpha = 2\sqrt{(a^\gamma)^2 - \frac{d_A^2}{4}}$. In order to reduce the misfit at the γ-α interface or at the interface between two α variants, we also assume that the dense direction $\mathbf{a}^\gamma$ remains invariant (untilted and undistorted). The invariance of a dense direction of type <100> or <110> is often observed for real cubic-tetragonal transformations, such as in AuCu alloys for example. An equivalent way to imagine the distortion would be to consider it as a stretch from the dashed green square in Figure 11b to the dashed red rectangle in Figure 11c, and then add a rotation in order to maintain the direction $\mathbf{a}^\gamma$ invariant. This compensating rotation is called "obliquity correction". Here, we decided to calculate directly the lattice distortion corresponding to the final orientation Figure 11c, and deduce the stretch and rotation by polar decomposition. Let us do the calculation with the supercell $\mathbf{B}_{super}^\gamma = 2\begin{pmatrix} 1 & 0 \\ 0 & 1 \end{pmatrix}$, $\mathbf{B}_{super}^{\gamma'} = 2\begin{pmatrix} 1 & Cos(\beta) \\ 0 & Sin(\beta) \end{pmatrix}$, and $\mathbf{B}_{super}^\alpha = \begin{pmatrix} 1 & 1 \\ -1 & 1 \end{pmatrix}$. The distortion matrix expressed in the basis associated with the supercell is given by equation (7): $\mathbf{F}_{super}^\gamma(\beta) = \mathbf{B}_{super}^{\gamma'} \left(\mathbf{B}_{super}^\gamma\right)^{-1} = \begin{pmatrix} 1 & Cos(\beta) \\ 0 & Sin(\beta) \end{pmatrix}$. Here, it is also the distortion matrix in the crystallographic basis $\mathbf{F}_c^\gamma$. Once the distortion is finished, the angular parameter $\beta_f$ is such that $Cos\left(\frac{\beta_f}{2}\right) = \frac{d_A}{2a^\gamma}$. This ratio will be called *r*. By using trigonometry, the distortion matrix is written $\mathbf{F}_c^\gamma = \begin{pmatrix} 1 & 2r^2 - 1 \\ 0 & 2r\sqrt{1-r^2} \end{pmatrix}$. It can be checked that if the diameter of the atoms A is such that $d_A = \sqrt{2}a^\gamma$, i.e. $r = \frac{1}{\sqrt{2}}$, then $\mathbf{F}_c^\gamma = \mathbf{I}$, there is no distortion. The orientation matrix is calculated with equation (8); it is $\mathbf{T}_c^{\gamma \to \alpha} = \begin{pmatrix} 2r^2 & -2(1-r^2) \\ 2r\sqrt{1-r^2} & 2r\sqrt{1-r^2} \end{pmatrix}$. The correspondence matrix is $\mathbf{C}_c^{\alpha\to\gamma} = (\mathbf{T}_c^{\gamma\to\alpha})^{-1}\mathbf{F}_c^\gamma = \frac{1}{2}\begin{pmatrix}1 & 1\\-1 & 1\end{pmatrix}$. By polar decomposition in an orthonormal basis, the distortion matrix is written $\mathbf{F}_c^\gamma = \mathbf{F}_\#^\gamma = \mathbf{Q}_\#^\gamma \mathbf{U}_\#^\gamma = \begin{pmatrix}\frac{r+\sqrt{1-r^2}}{\sqrt{2}} & \frac{r-\sqrt{1-r^2}}{\sqrt{2}}\\ \frac{-r+\sqrt{1-r^2}}{\sqrt{2}} & \frac{r+\sqrt{1-r^2}}{\sqrt{2}}\end{pmatrix}\begin{pmatrix}\frac{r+\sqrt{1-r^2}}{\sqrt{2}} & \frac{r-\sqrt{1-r^2}}{\sqrt{2}}\\ \frac{r-\sqrt{1-r^2}}{\sqrt{2}} & \frac{r+\sqrt{1-r^2}}{\sqrt{2}}\end{pmatrix}$. The eigenvalues of the stretch matrix give the contraction and elongation values; they are $\sqrt{2}\,r$ and $\sqrt{2}\sqrt{1-r^2}$. The obliquity angle $\xi$ is given by the rotation angle of $\mathbf{Q}_\#^\gamma$; it is $\xi = ArcCos(\frac{r+\sqrt{1-r^2}}{\sqrt{2}})$. One can check that $\xi = 0$ for $r = \frac{1}{\sqrt{2}}$. The calculations also show that $\mathbb{H}_F^\gamma = \mathbb{H}_D^\gamma = \mathbb{H}_T^\gamma = \{g_1^{Sq}, g_4^{Sq}\}$ and $\mathbb{H}_C^\gamma = \{g_1^{Sq}, g_4^{Sq}, g_6^{Sq}, g_7^{Sq}\}$, thus $N_F^\alpha = N_D^\alpha = N_T^\alpha = 4$ and $N_C^\alpha = 2$. This example is good prototype of order-disorder cubic-tetragonal displacive transformation. The fcc-bct (body-centred tetragonal) martensitic transformation in steels also implies an ordering of the interstitial carbon atoms, but the kinetics of carbon diffusion is so high that martensitic transformation in steels is often very fast and athermal, whereas the order-disorder displacive transformations are generally slow and activated by thermal treatments (as in AuCu alloys).

### 7.9. Cycle of transformations Square *p*4*mm* ↔ Triangle *p*3*m*1

Let us consider again the case of Rectangle *p*4*mm* → Triangle *p*3*m*1 treated in § 7.4, and its inverse transformation Triangle *p*3*m*1→ *p*4*mm*, which is very similar to the example in § 7.6. The effect of cycling of Square ↔ Triangle transformations on the orientation variants can be explicitly shown by the squares and triangles formed at each generation (Figure 12a). The γ→α and the α→γ misorientations can be schematized by segments forming a graph (Figure 12b). In this example, all the orientation variants created at the second generation were already created at an earlier generation (here 1st generation), which implies that the graph is finite, i.e. $\gamma^1\to\{\alpha^1\}\to\{\gamma^2\}\to\{\alpha^2\} \equiv \{\alpha^1\}$. This can be proved by encoding a square and a triangle by the angles of their edges with the horizontal (at 0 rad.), i.e. by the congruence classes [0] # π/2 and [0] # 2π/3, respectively. Note that the triangle obtained by horizontal reflection is [π/3] # 2π/3. The triangles of first generation are encoded by the set {[0],[π/3], [π/2], [-π/2]} # 2π/3. The squares of second generation form the set {[π/3],[-π/3]} # π/2. The triangles of second generation are obtained by adding the two previous sets; by using the properties of the "modulo" operation, one can show that no new triangle variant is created after the second cycle. The total number of triangle and square variants are $N^{(\alpha^1+\alpha^2)} = 4$ and $N^{(\gamma^1+\gamma^2)} = 3$, respectively. The orientation variants and their graph are shown in Figure 12c,d. The graph is finite because of the special square-triangle orientation relationship and because of the rationality between the angles π/3 and π/2.

**7.10. Cycle of transformations Triangle *p*3*m*1 ↔ Triangle *m***

Let us now consider the case of a Triangle *p*3*m* (γ) ↔ Triangle *m* (α) transformations shown in Figure 13. The triangle *m* is isosceles and chosen such that the angle between the two sides of equal length is ξ = ArcCos(1/3). Niven's theorem states that for any angle $\theta = q\pi, q \in \mathbb{Q}$, if $Cos(\theta) \in \mathbb{Q}$, then $Cos(\theta) = 0, \frac{1}{2}$ or 1, i.e. $\theta$ = π/2, π/3 or 0. Consequently, the ratio of ξ (the angle of the isosceles triangle) divided by π/3 (the angle of equilateral triangle) is not rational; which implies that there is no possibility to reobtain the same initial orientation whatever the number of cycles of transformation. The orientation variants can be represented by a fractal (only a part, arbitrarily stopped at the second generation, is shown in Figure 13a), and the orientation graph is infinite. This 2D example helps to understand why the orientation graph associated with the fcc-bcc transformation with a KS OR is also infinite. Indeed, the KS OR implies a parallelism of the planes (111)$_\gamma$ // (110)$_\alpha$. The (111)$_\gamma$ plane can be represented by the triangle *p*3*m* formed by the 3 dense directions <110>$_\gamma$ it contains, and the (110)$_\alpha$ plane by the triangle *pm* of angle ξ = ArcCos(1/3) formed by the two dense direction <111>$_\alpha$ it contains, as illustarted in Figure 13b. The symmetries of the (110)$_\alpha$ plane is actually *pmm* because the rectangle is made with the [001]$_\alpha$ and [$\bar{1}$10]$_\alpha$ directions, but this does not affect the fact that the number of variants sharing the same dense plane (111)$_\gamma$ // (110)$_\alpha$ become infinite by increasing the number of cycles.

**8. Conclusions**

This paper recalls the importance of distinguishing three types of transformation matrices to describe a displacive phase transformation γ→α: the lattice distortion matrix $\mathbf{F}_c^\gamma$, the orientation relationship matrix $\mathbf{T}_c^{\gamma\to\alpha}$, and the correspondence matrix $\mathbf{C}_c^{\gamma\to\alpha}$. The three matrices are linked by the relation $\mathbf{C}_c^{\alpha\to\gamma} = \mathbf{T}_c^{\alpha\to\gamma}\mathbf{F}_c^\gamma$, with $\mathbf{T}_c^{\alpha\to\gamma}$ and $\mathbf{C}_c^{\alpha\to\gamma}$ the inverse matrices of $\mathbf{T}_c^{\gamma\to\alpha}$ and $\mathbf{C}_c^{\gamma\to\alpha}$, respectively. The index "c" means that the matrices are expressed in their natural crystallographic basis. They can be written in an orthonormal basis thanks to the structure tensor, and in the reciprocal basis thanks to the metric tensor. The stretch matrix $\mathbf{U}_c^\gamma$ is deduced from the distortion matrix $\mathbf{F}_c^\gamma$ by usual polar decomposition once the matrix is written in an orthonormal basis, i.e. by calculating $\left(\mathbf{F}_\#^\gamma\right)^t \mathbf{F}_\#^\gamma$, or by a specific polar decomposition that respects the metrics, i.e. by calculating $\left(\mathbf{F}_c^\gamma\right)^t \mathcal{M}^\gamma \mathbf{F}_c^\gamma$. The transformation matrices for the inverse transformation α→γ can be deduced from those of the direct transformation γ→α thanks to the matrices $\mathbf{T}_c^{\gamma\to\alpha}$ and $\mathbf{C}_c^{\gamma\to\alpha}$ as described in § 2. The three transformation matrices can be explicitly calculated from the supercell used to establish the correspondence between the two phases, as detailed in § 3. The distortion matrix $\mathbf{F}_c^\gamma$ is generally determined for the complete transformation between initial state (parent phase) and final state (daughter phase), but in the cases where the atoms behave as hard-spheres of constant size, it is possible to establish a continuous analytical form $\mathbf{F}_c^\gamma(\theta)$ for the distortion that depends on a unique angular parameter $\theta$ varying between a start value $\theta_s$ and a final value $\theta_f$ such that $\mathbf{F}_c^\gamma(\theta_s) = \mathbf{I}$ and $\mathbf{F}_c^\gamma(\theta_f) = \mathbf{F}_c^\gamma$. The continuous form is not a simple strain due to the steric effect, and Schmid's law cannot be applied. New criteria that aim at predicting the formation of the daughter phase or deformation twins under stress can be proposed based on $\mathbf{F}_c^\gamma(\theta)$ or on its derivative given by the formula $\frac{D\mathbf{F}}{D\theta} = \frac{d\mathbf{F}}{d\theta} \mathbf{F}^{-1}$, as introduced in § 4.

Variants are defined for each of the three types of transformation matrix. They are the cosets built on an intersection group $\mathbb{H}^\gamma$ that is a subgroup of the point group of the parent phase $\mathbb{G}^\gamma$. More precisely, the intersection groups for the distortion, orientation, and correspondence variants are $\mathbb{H}_F^\gamma = \mathbb{G}^\gamma \cap \mathbf{F}_c^\gamma \mathbb{G}^\gamma (\mathbf{F}_c^\gamma)^{-1}$, $\mathbb{H}_T^\gamma = \mathbb{G}^\gamma \cap \mathbf{T}_c^{\gamma \to \alpha} \mathbb{G}^\alpha (\mathbf{T}_c^{\gamma \to \alpha})^{-1}$, and $\mathbb{H}_C^\gamma = \mathbb{G}^\gamma \cap \mathbf{C}_c^{\gamma \to \alpha} \mathbb{G}^\alpha (\mathbf{C}_c^{\gamma \to \alpha})^{-1}$, respectively. The stretch variants, sometimes improperly called "correspondence" variants, are actually like distortion variants; they are defined as cosets built on $\mathbb{H}_U^\gamma = \mathbb{G}^\gamma \cap \mathbf{U}_c^\gamma \mathbb{G}^\gamma (\mathbf{U}_c^\gamma)^{-1}$. The number of variants is directly given by Lagrange's formula whatever their type. As shown in § 5, the number of variants of direct transformation is related to the number of variants of reverse transformation by the equation $N^\alpha |\mathbb{G}^\alpha| = N^\gamma |\mathbb{G}^\gamma|$, where $|\mathbb{G}^\alpha|$ and $|\mathbb{G}^\gamma|$ are the orders of the point groups of α and γ phases, respectively. The difference between the orientation and correspondence variants is subtle but important from a theoretical point of view because it implies a distinction between the geometric symmetries and the algebraic symmetries. Two equivalent geometric symmetries are defined by the same geometric element (rotation, reflection etc.), but they can be written with different matrices according to the crystal in which the symmetry operates. Two equivalent algebraic symmetries are defined by the same matrix, and thus same permutation of basis vectors, independently of the crystal in which they operate.
Reversibility/irreversibility during phase transformation cycling is a multifactor property. One of its aspects, the orientation reversibility is briefly introduced in § 6 by expressing variants of *n*-generation with *n*-cosets and by representing them with fractal graphs. Numerous 2D examples are given in § 7 in order to show that there is no general relation between the numbers of distortion, orientation variants and correspondence variants, and to illustrate the concept of orientation variants formed during thermal cycling. The subject of *n*-variants generated by cycling and the related reversibility/irreversibility issue remain widely open for future research.

The three transformation matrices are related to different characteristics of phase transformation; the orientation variants are of interest for the microstructural observations (EBSD, X-ray diffraction), the distortion variants for the mechanical (transformation induced plasticity and shape-memory) properties, and the correspondence variants for the change of the atomic bonds (spectrometry). We hope that the mathematical definitions and equations proposed in this work will be useful for future research in these fields.

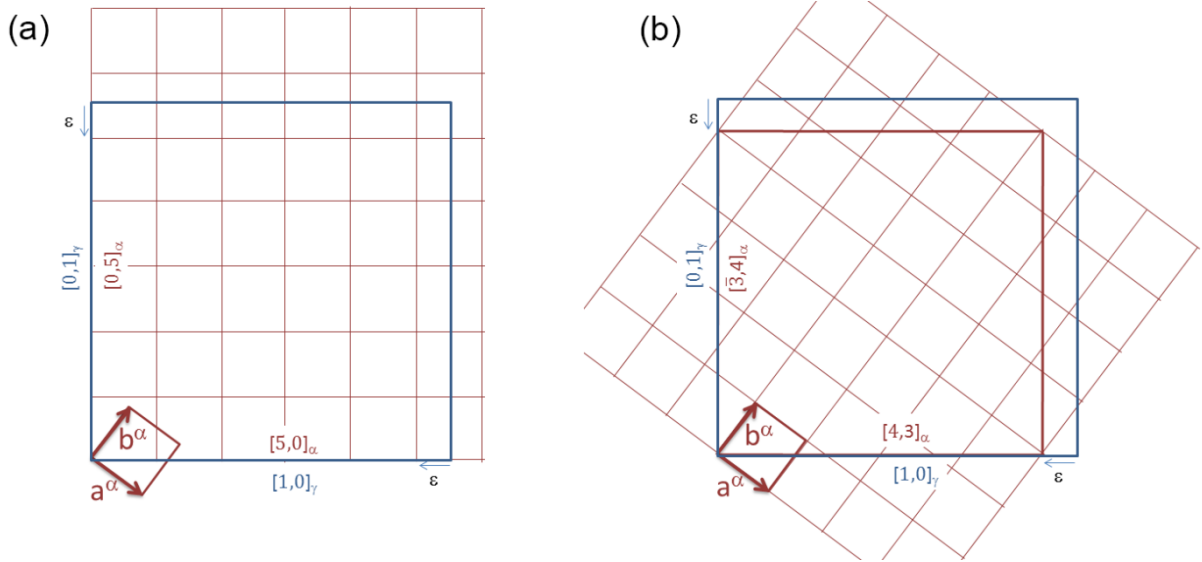

**Figure 1** Difference between the stretch matrix and the correspondence matrix illustrated with a square γ - square α transformation with two ORs. (a) <1,0>γ // <5,0>α, and (b) <1,0>γ // <4,3>α. The lattice parameter of the square γ is nearly five times that of square α. The distortion (stretch) matrix is the same in both cases; it is a diagonal matrix $\begin{pmatrix} r & 0 \\ 0 & r \end{pmatrix}$ with $r = 1 + \varepsilon = \frac{a_\gamma}{5a_\alpha}$, but the orientation and correspondence matrices are different.

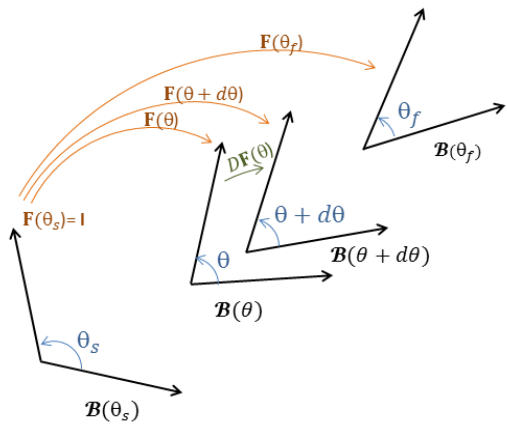

**Figure 2** Derivative of a continuous distortion matrix $\mathbf{F}(\theta)$. The parameters $\theta_s$ and $\theta_s$ are associated with start and finish states of the transformation. The matrix $\mathbf{F}(\theta)$ is defined as a coordinate transformation matrix from the starting basis $\mathcal{B}(\theta_s)$ to the basis $\mathcal{B}(\theta)$, i.e. $\mathbf{F}(\theta) = [\mathcal{B}(\theta_s) \rightarrow \mathcal{B}(\theta)]$. The infinitesimal matrix at the state $\theta$ expressed in the local basis $\mathcal{B}(\theta)$ is
$D\mathbf{F}(\theta)_{loc} = [\mathcal{B}(\theta) \rightarrow \mathcal{B}(\theta + d\theta)] = \mathbf{F}(\theta)^{-1}\mathbf{F}(\theta + d\theta)$.

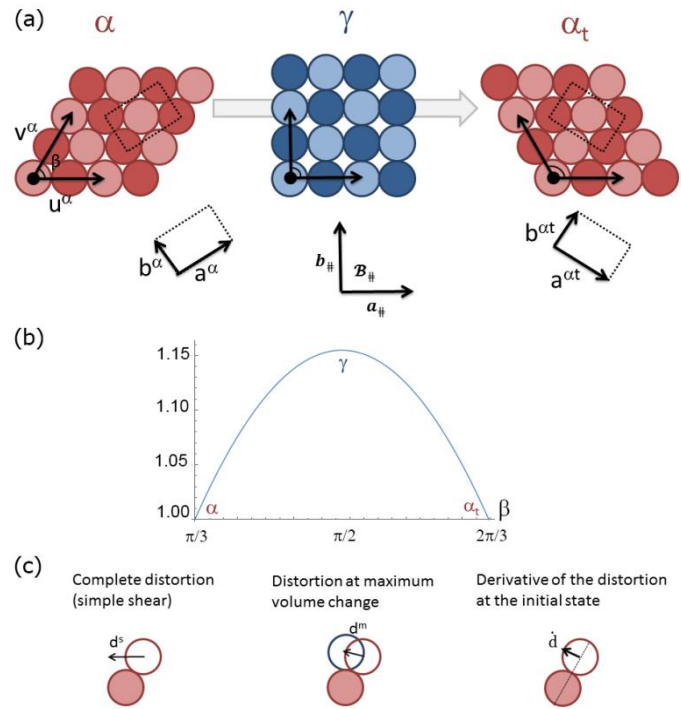

**Figure 3** 2D simple case of continuous deformation twinning of a p2mm α phase made of two atoms considered as hard-spheres (disks) of same constant size. (a) Schematic view of the distortion between the initial crystal α and its twin $α_t$. A cubic γ appears as an intermediate state. The angular parameter β is $β_s = π/3$ in the initial state, $β_i = π/2$ in the intermediate cubic state, and $β_f = 2π/3$ in the final state. (b) Volume change during the distortion. (c) Three hypotheses can be imagined to calculate the interaction work for a twinning criterion: (i) with the complete distortion $\mathbf{F}^α(β_f)$, which is here a simple strain, (ii) with the distortion at the maximum volume change $\mathbf{F}^α(β_i)$, or (iii) with the derivative of the distortion at the starting state $\frac{D\mathbf{F}}{D\beta}(β_s)$.

October 2018

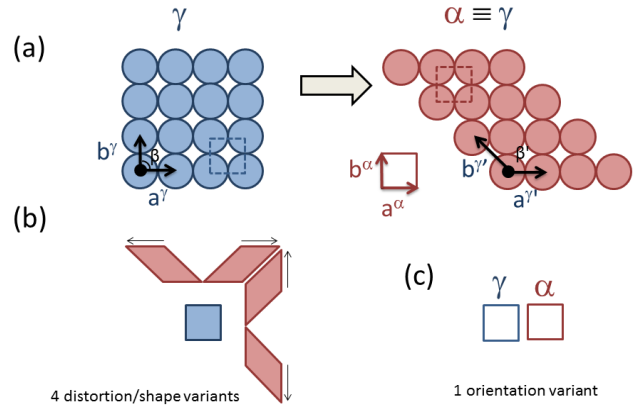

**Figure 4** Square p4mm ($\gamma$) $\rightarrow$ Square p4mm ($\alpha$) transformation with $\Sigma 1$ orientation by simple strain. (a) The lattice and its "atoms" before distortion (in blue) and after distortion (in salmon). (b) Four distortion variants, (c) one orientation variant, and four correspondence variants (not represented) are generated.

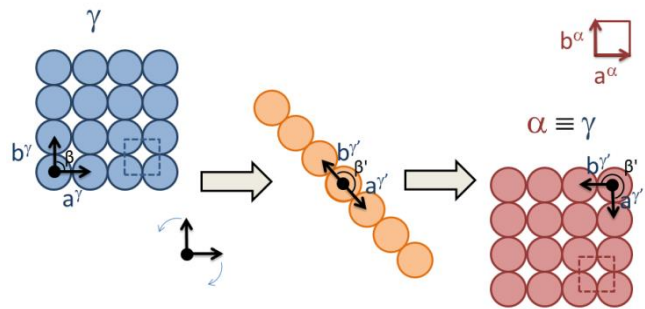

**Figure 5** Square p4mm ($\gamma$) $\rightarrow$ Square p4mm ($\alpha$) transformation with $\Sigma 1$ orientation by turning inside-out. The angle $\beta$ changes continuously from $\beta_s = \pi/2$ to $\beta_f = 3\pi/2$. Two distortion variants, one shape-distorted variant, one orientation variant, and one correspondence variant are generated (they are not represented). The determinant of the distortion matrices at the different stages are Det($\mathbf{F}^\gamma (\beta_s)$) > 0, Det($\mathbf{F}^\gamma (\beta_f)$) < 0, and Det($\mathbf{F}^\gamma (\beta_i)$) = 0. This example show a mathematical case where the distortion and the distortion-shape variants are distinct; however, the distortion is considered as "non-crystallographic" because it implies an intermediate degenerated state at $\beta_i = \pi$.

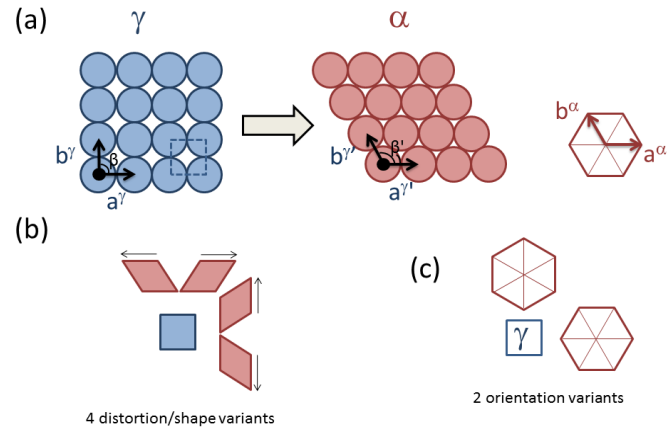

**Figure 6** Square p4mm ($\gamma$) $\rightarrow$ Hexagon p6mm ($\alpha$) transformation by angular distortion. (a) The lattice and its "atoms" before distortion (in blue) and after distortion (in salmon). (b) Four distortion variants, (c) two orientation variants, and two correspondence variants (not represented) are generated.

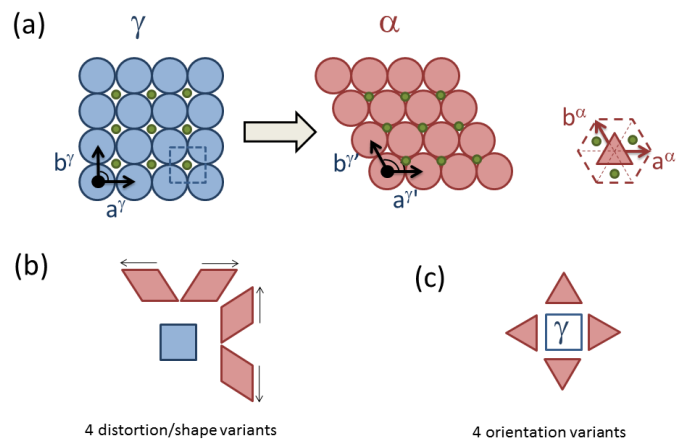

**Figure 7** Square p4mm ($\gamma$) $\rightarrow$ Triangle p3m1 ($\alpha$) transformation by angular distortion. The parent phase is constituted of large atoms (in blue) and small interstitial atoms (in green). (a) The lattice and its "atoms" before distortion (in blue) and after distortion (in salmon). The interstitial atoms must choose between two equivalent sets of positions; their trajectories do not follow the lattice distortion (they "shuffle"). (b) Four distortion variants, (c) four orientation variants, and four correspondence variants (not represented) are generated.

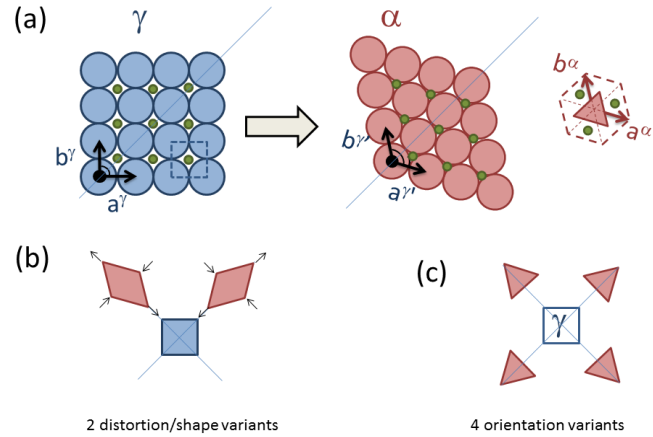

**Figure 8** Square p4mm (γ) → Triangle p3m1 (α) transformation by pure stretching. The parent and daughter phases are the same as in the previous example. (a) The lattice and its "atoms" before distortion (in blue) and after distortion (in salmon). (b) Two distortion variants, (c) four orientation variants, and two correspondence variants (not represented) are generated. Note that there are more orientation variants than distortion variants.

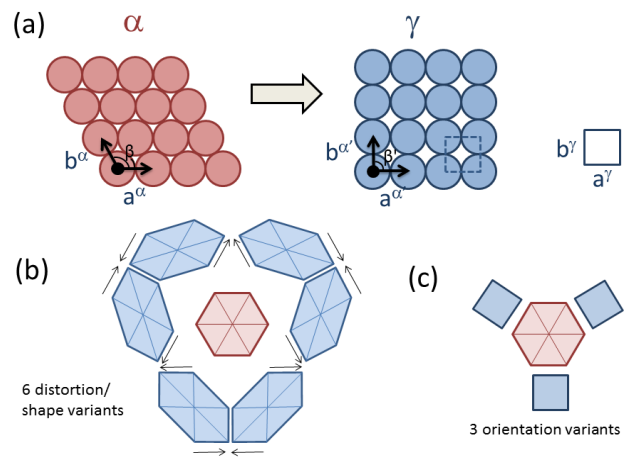

**Figure 9** Hexagon p6mm (α) → Square p4mm (γ) transformation by angular distortion. This is the reverse transformation of the example shown in Figure 6. (a) The lattice and its "atoms" before distortion (in salmon) and after distortion (in blue). (b) Six distortion variants, (c) three orientation variants, and three correspondence variants (not represented) are generated.

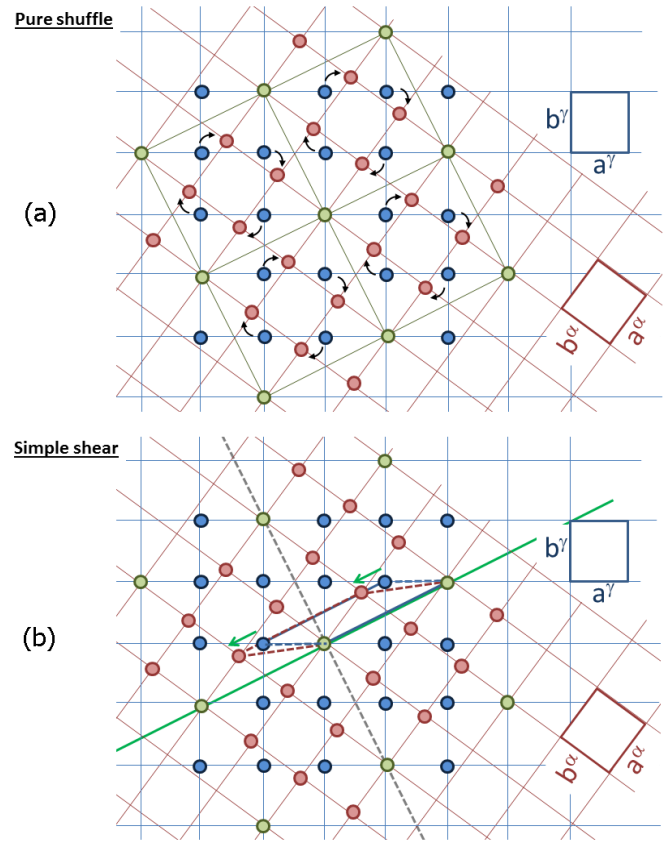

**Figure 10**   Square p4mm (γ) → Square p4mm (α) transformations with Σ5 OR implying different mechanisms: (a) pure shuffle (curves arrows) and (b) simple strain (green arrows). In case (a) the atoms at the nodes of the Σ5 CSL (in light green) do not move, whereas they move by five times the green arrow in case (b).

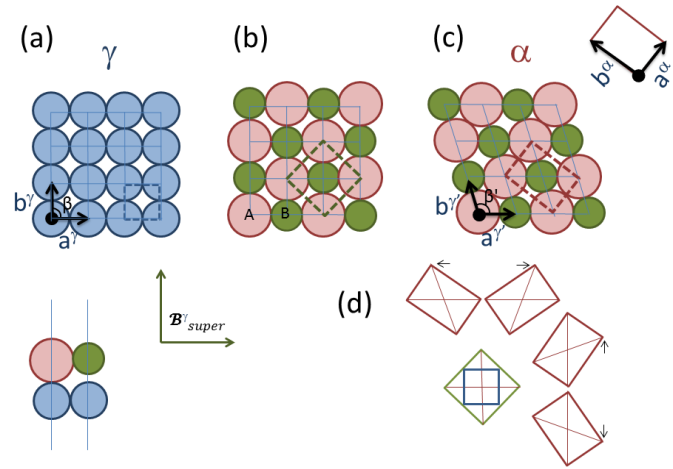

**Figure 11** Square p4mm ($\gamma$) $\rightarrow$ Rectangle p2mm ($\alpha$) transformation with an order/disorder displacive character. The lattice is constituted of two different atoms considered as hard-spheres of different size: the large atoms A in salmon and the small atoms B in green. (a) At high temperature, in the disordered state, the lattice nodes are randomly occupied by A or B; the "mean" atom is coloured in blue. At lower temperature, the atoms organize themselves depending on their affinities. (b) If only ordering is considered without lattice distortion, the ordered state remains cubic. (c) If the atoms A come in contact to each other, the ordered phase becomes rectangular. (d) Four distortion variants, four orientation variants, and two correspondence variants (not represented) are generated.

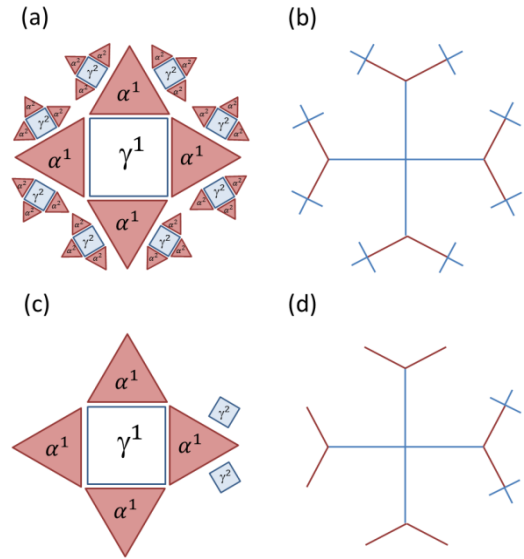

**Figure 12** Cycling of Square p4mm ($\gamma$) ↔ Triangle p3m1 ($\alpha$) transformations. (a) Fractal representation of the orientation variants. The upper indices $n$ represent the $n^{th}$ generation. The variant lower indices are not reported for sake of simplicity. (b) Graph representation, with the blue and salmon segments representing the Square → Triangle and the Triangle → Square transformations, respectively. After removing the redundant orientation variants, (a) and (b) are simplified into (c) and (d) respectively: the cycling graph is finite.

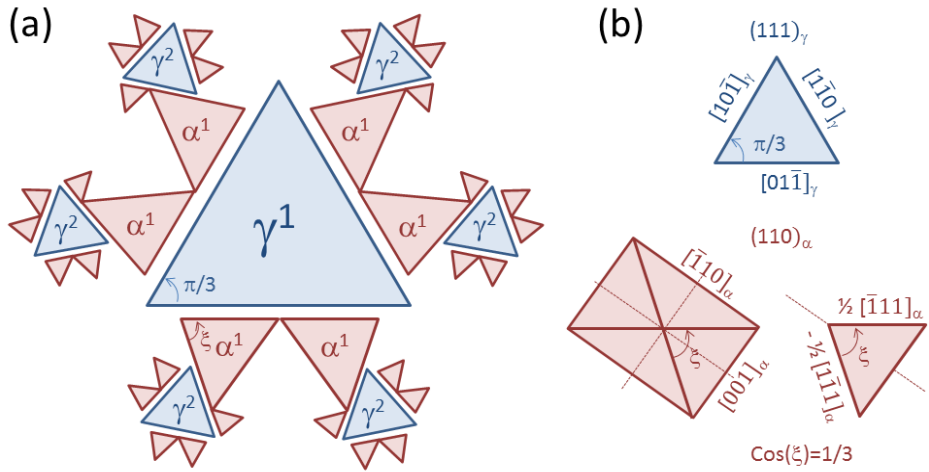

**Figure 13** Cycling of Triangle p3m1 ($\gamma$) ↔ Triangle pm ($\alpha$) transformations. The triangle pm is isosceles and chosen such that the angle between the two sides of equal length is $\xi$ = ArcCos(1/3). (a) Fractal representation of the orientation variants (arbitrarily stopped at the second generation). (b) Equivalence between this 2D example and a part of the 3D KS OR between fcc ($\gamma$) and bcc ($\alpha$) phases. The equilateral triangle p3m1 is formed by the 3 dense directions $<110>_\gamma$ in the plane $(111)_\gamma$. The isosceles triangle pm is formed by the two dense direction $<111>_\alpha$ in the $(110)_\alpha$ plane. The fractal and its associated cycling graph (not shown here) are infinite (the proof is given by Niven's theorem).

**Table 1** Different type of variants created by structural phase transformations. They are defined as cosets based on subgroups of the point group $\mathbb{G}^\gamma$ of the parent phase.

| Variants | Subgroup of $\mathbb{G}^\gamma$ | Number of variants |
|---|---|---|
| Orientation | $\mathbb{H}_T^\gamma = \mathbb{G}^\gamma \cap \mathbf{T}_c^{\gamma\to\alpha} \mathbb{G}^\alpha (\mathbf{T}_c^{\gamma\to\alpha})^{-1}$ | $N_T^\alpha = \dfrac{|\mathbb{G}^\gamma|}{|\mathbb{H}_T^\gamma|}$ |
| Correspondence | $\mathbb{H}_C^\gamma = \mathbb{G}^\gamma \cap \mathbf{C}_c^{\gamma\to\alpha} \mathbb{G}^\alpha (\mathbf{C}_c^{\gamma\to\alpha})^{-1}$ | $N_C^\alpha = \dfrac{|\mathbb{G}^\gamma|}{|\mathbb{H}_C^\gamma|}$ |
| Distortion | $\mathbb{H}_F^\gamma = \{g_i^\gamma \in \mathbb{G}^\gamma,\ g_i^\gamma \mathbf{F}_c^\gamma (g_i^\gamma)^{-1} = \mathbf{F}_c^\gamma\}$ | $N_F^\alpha = \dfrac{|\mathbb{G}^\gamma|}{|\mathbb{H}_F^\gamma|}$ |
| Distorted-Shape | $\mathbb{H}_D^\gamma = \mathbb{G}^\gamma \cap \mathbf{F}_c^\gamma \mathbb{G}^\gamma (\mathbf{F}_c^\gamma)^{-1}$ | $N_D^\alpha = \dfrac{|\mathbb{G}^\gamma|}{|\mathbb{H}_D^\gamma|}$ |
| Stretch | $\mathbb{H}_U^\gamma = \{g_i^\gamma \in \mathbb{G}^\gamma,\ g_i^\gamma \mathbf{U}_c^\gamma (g_i^\gamma)^{-1} = \mathbf{U}_c^\gamma\}$ | $N_U^\alpha = \dfrac{|\mathbb{G}^\gamma|}{|\mathbb{H}_U^\gamma|}$ |

**Table 2** Number of distortion variants $N_F^\alpha$, orientation variants $N_T^\alpha$, and correspondence variants $N_C^\alpha$ in the different examples treated in §7. The number of shape-distortion variants $N_D^\alpha$ is not reported because it always the same as the number of distortion variants (excepted in the "non-crystallographic" example of §7.2).

| Section | $N_F^\alpha$ | $N_T^\alpha$ | $N_C^\alpha$ | Figure |
|---|---|---|---|---|
| §7.1 | 4 | 1 | 4 | Figure 4 |
| §7.2 | 2 | 1 | 1 | Figure 5 |
| §7.3 | 4 | 2 | 2 | Figure 6 |
| §7.4 | 4 | 4 | 4 | Figure 7 |
| §7.5 | 2 | 4 | 2 | Figure 8 |
| §7.6 | 6 | 3 | 3 | Figure 9 |
| §7.7.1 | 1 | 2 | 2 | Figure 10a |
| §7.7.2 | 4 | 2 | 4 | Figure 10b |
| §7.8 | 4 | 4 | 2 | Figure 11 |

**Acknowledgements**   We acknowledge Prof. Roland Logé, director of the EPFL LMTM laboratory, and Dr Andreas Blatter, scientific director of PX Group, for their strong support and help.

**Appendix A. The phenomenological theory of martensitic transformation (PTMC) and its possible alternative**

**A1. From Bain's distortion to PTMC**

A well-known lattice distortion between fcc (γ) austenite and bcc (α) martensite in steels was proposed by Bain (1924) who noticed that a fcc crystal could be transformed into a bcc crystal by a contraction along a <001>$_\gamma$ axis and an extension along the <110>$_\gamma$ axes perpendicular to the contraction axis. Bain distortion **U** (also often noted **B**) is the prototype of stretch fcc-bcc distortion. Let us recall that a stretch matrix is a symmetric matrix; it is diagonal in an appropriate orthonormal basis. Any distortion can be written by polar decomposition into a combination of rotation and stretch. From Bain's model, one should expect that the parent/daughter orientation relationship is a rotation of π/4 around the <100>$_\gamma$ axes. However, Kurdjumov & Sachs (KS) (1930) and Nishiyama (1934) found by X-ray diffraction ORs in which the dense planes are (111)$_\gamma$ // (110)$_\alpha$, with, for KS OR, a parallelism of the dense directions [$\bar{1}$10]$_\gamma$ // [$\bar{1}$11]$_\alpha$, or for Nishiyama a parallelism [$\bar{1}$10]$_\gamma$ // [001]$_\alpha$. The two ORs are quite close (5.2°) but both are 10° far from that expected from Bain's model. This discrepancy made Kurdjumov, Sachs and Nishiyama propose another model in which the transformation occurs by a combination of shear and stretch. The model was discarded by Greninger & Troiano (1949) because of the high value of the shear amplitude. In the same paper, Greninger & Troiano proposed new ideas based on shear composition to build a crystallographic model that gave birth to the phenomenological theory of martensitic transformation (PTMC). It is assumed in PTMC that the shape of martensite follows an invariant plane strain (IPS). An IPS is a matrix of type **I** + **d** · **p**$^t$, where **I** is the identity matrix, **d** is the shear direction (in column) and **p** is the shear plane (with **p**$^t$ in line), which is also the plane of interface between the martensite product and the austenite matrix according to PTMC. One can check that any vector belonging to the plane **p** (i.e. normal to the vector **p**) is invariant, and that any vector along **p** is strained by a factor 1+ **d**$^t$. **p**. An IPS is a simple strain when there is no volume change (**d**$^t$. **p** = 0). Bowles & Mackenzie (1954), independently and nearly with simultaneously with Wechsler, Lieberman & Read (1953), introduced with many details the PTMC, following Greninger & Troiano's ideas. Bowles & Mackenzie's version of PTMC is based on the assumption that the final shape **S** should be an IPS and that this IPS should be the combination of the Bain distortion **U**, with an additional rotation **R** (a free parameter that should fit the equations), and an inhomogeneous lattice invariant shear **P** (twinning or slip system), following an equation of type **S** = **Q**.**U**.**P**. Its input parameters are the lattice parameters of the parent and daughter phases (from which **U** is deduced). The fitting parameters are the choice of the LIS system for **P** and the amplitude of shear. The output parameters are the OR (given by the additional rotation **Q**) and the IPS **S**, and thus the habit plane of martensite product. An important step in solving the equations consists in noticing that **S**. **P**$^{-1}$ = **Q**.**U** should have invariant line (given by the intersection of the two invariant planes associated with **S** and **P**). Note that the martensite product is not homogeneous but contains series of defects or twins whose amount depend on the shear amplitude of **P**. The shear **P** imagined in this model is purely mechanical; it does not lead to a new variant if **P** is a slip, or it can lead to a new orientation if **P** is a twin. Also note that the equation only implies matrix multiplications, which means that the operations are combined successively and not simultaneously. This is a problem because matrix product is non-commutative in general, which means that the result depends on the choice of the order of the operations. To solve this problem, infinitesimal version of the PTMC were intended with the Bowles & Mackenzie formalism, without success (Kelly, 2003).

Wechsler, Lieberman & Read's formalism is quite different. It has been continued to be developed for shape memory alloys (SMA) by Ball & James (1987), James & Hane (2000), and Battacharya (2003) by using high-level mathematics and by generalizing the concept of kinematic compatibility for groups of martensite variants. This version of PTMC considers that the shape deformation **S** results from the association of two variants in volume faction $x$ and (1-$x$), i.e. $\mathbf{S} = x\mathbf{F}_1 + (1-x)\mathbf{F}_2$ where $\mathbf{F}_1$ and $\mathbf{F}_2$ are the distortions associated to variants 1 and 2, respectively. The martensite product is imagined as a composite made of $x$ variant 1 and (1-$x$) variant 2 in volume proportion. In order to get the expected atomic volume change, Det(**S**) should be equal Det(**F**) = vol(bcc)/vol(fcc), which is quite restrictive because the determinant is not a linear function of matrices; the equality is obtained if and only if there is continuity (compatibility) of the interface between the variants 1 and 2, i.e. the interface on the plane $\mathbf{p}_{21}$ is distorted similarly by $\mathbf{F}_1$ and $\mathbf{F}_2$ (it can be rotated). Mathematically this condition is equivalent to $\mathbf{F}_1 - \mathbf{F}_2 = \mathbf{d}_{21} \cdot \mathbf{p}_{21}^t$. This equation can also be written $\mathbf{F}_1 = \mathbf{F}_2 \cdot (\mathbf{I} + \mathbf{F}_2^{-1} \mathbf{d}_{21} \cdot \mathbf{p}_{21}^t) = \mathbf{F}_2 \mathbf{P}_{21}$, with $\mathbf{P}_{21} = \mathbf{I} + (\mathbf{F}_2^{-1} \mathbf{d}_{21}) \cdot \mathbf{p}_{21}^t$, which shows that the variant 2 can be transformed into the variant 1 by a simple strain $\mathbf{P}_{21}$ on the plane $\mathbf{p}_{21}$ along the direction by $\mathbf{F}_2^{-1} \mathbf{d}_{21}$. This fact justifies that the two variants generated by a phase transformation are also mechanical twins (if the shear amplitude is realistic), and that variant reorientation in SMA can be obtained by deformation. The compatibility criterion between the austenite and the martensite product is equivalent to say that **S** is an IPS. In addition, by using $\mathbf{F}_1 = \mathbf{F}_2 \mathbf{P}_{21}$ and writing $\mathbf{P}_{21} = x\mathbf{P}$ and $\mathbf{F}_2 = \mathbf{QU}$ by polar decomposition, one can easily check that the equation $\mathbf{S} = x\mathbf{F}_1 + (1-x)\mathbf{F}_2$ is also written **S** = **Q**.**U**.**P**. Consequently, Wechsler, Lieberman & Read's theory is equivalent to Bowles and Mackenzie's one, despite the apparent difference of formalism, which justifies the unique name "PTMC" given to both versions of the theory. The compatibility conditions are sensitive to the lattice parameters. Both versions explicitly or implicitly use the Bain correspondence matrix **C** that was first proposed by Jaswon & Wheeler (1948). The distortion matrix **F** and the orientation relationship matrix **T** are indeed linked together by **C**, according to **F** = **T**.**C** (see also § 2.3). This relation was explicitly used by Bowles & Mackenzie (1954) and in the crystallographic theories of deformation twinning (see for example Bevis & Crocker, 1968; Christian & Mahajan, 1995; Szczerba *et al*. 2012).

**A2. Some compatibility criteria used in PTMC**

The understanding of the formation of variants is of importance to get a better understanding of the reversibility of martensitic alloys (Bhattacharya *et al*., 2004; Cayron, 2006; Gao *et al*., 2017). One may distinguish the reversibility of the stress-induced transformations evaluated on SMA by tensile straining tests, from the reversibility of temperature-induced transformations evaluated by thermal cycles between temperatures below $M_f$ (martensite finish temperature) and above $A_f$ (austenite finish temperature). Different factors play a crucial role on temperature-induced reversibility. PTMC mainly treats interface reversibility, as it defines some supercompatibility criteria requiring on some relations between the lattice parameters of the parent and daughter phases, i.e. *i*) the compatibility between the parent and the martensite product, *ii*) the compatibility between the two variants forming the martensite product, as explained previously, and *iii*) a criterion called "cofactor condition" (Chen *et al*., 2013). Only the stretch variants $\mathbf{U}_i$ are required in the calculations. Beside these crystallographic considerations, the mechanical properties of the parent and daughter phases, mainly their yield strengths, are also important. Indeed, high yield strengths (or low transformation temperatures) generally implies a low activity of dislocations, which favours elastic accommodations or accommodation mechanisms without the usual plastic deformation modes (variants pairing or grouping). Hardening the parent phase by precipitation or by modifying the grain sizes helps for increasing the reversibility of SMA (Gu *et al*., 2018). For the last years, PTMC has tried to incorporate plasticity from pure crystallography. The set of accumulated deformation (and defects) during thermal cycling was identified to a group called "global group" by Bhattacharya *et al*. (2004). This group is built with two types of generators, the usual generators of the point group (the symmetries), and a primitive LIS (Eriksen, 1989). For example, the global group of a simple square "phase" is made of the LIS matrices generated by the simple strain $\begin{pmatrix} 1 & -1 \\ 0 & 1 \end{pmatrix}$ and by the square symmetries; it is the special linear group $SL(2, \mathbb{Z})$ of matrices of determinant 1 made of integer coefficients. This approach has been followed by Gao *et al*. (2017) who introduced geometric representations (Cayley graphs) of the global group of lattice deformation. We are not yet fully convinced by this approach in its present form because plasticity seems more complex than combinations of symmetries and LIS. It is not clear from a mechanical point of view if the usual way to write space group operations combining orientation symmetries *g* and translation symmetries *t* with the Seitz symbols {*g* | *t*} and their associated semi-direct product can be transformed into "global" shear matrices with their usual matrix product. Can we really assume that the translation operation {**E** | [1,0]} is equivalent to a simple strain operation $\begin{pmatrix} 1 & 1 \\ 0 & 1 \end{pmatrix}$?

## A3. An alternative approach to the PTMC

The success of PTMC is more significant and impressive for shape memory alloys (SMA) than for martensite in steels for which some habit planes such as {225} have remained unexplained for a long time (see for example Dunne & Wayman, 1971). For the last decade, we have tried to explain specific continuous features that appear in the pole figures of martensite in steels and other alloys, and we could not get any answer from PTMC. We thus came to conclude that the difficulty of PTMC for steels comes from the important distortion associated with the fcc-bcc transformation, which implies accommodation by dislocations and makes the incompatibility criteria of PTMC less relevant, or at least relaxed. Instead of using the exact values of the lattice parameters and guessing the LIS systems to fit the proportion of twins or LIS in order to reach an IPS, as practised in PTMC, we consider the OR as known, and we explain the continuous features observed in EBSD from the atomic displacements and lattice distortion of the transformation. Considering the OR as an input data is exactly the approach proposed by Nishiyama (1972) to overcome the failures of PTMC for steels. We introduced a type of distortion called "angular-distortive", and we imagine that martensite is formed from austenite in a way similar to a soliton wave; the OR is imposed by a condition on the wave propagation more than by the exact values of the lattices parameters (Cayron, 2018). We started by a two-step model implying an intermediate hexagonal phase (Cayron *et al.*, 2010), then a one-step model based on Pitsch OR (Cayron, 2013), and more recently a continuous one-step model based on KS OR (Cayron, 2015), generalized to other martensitic transformations between fcc, hcp and bcc phases (Cayron, 2016). Extension and contraction twinning modes in magnesium were also treated (Cayron, 2017a,b). In these works, the atoms were assumed to be hard-spheres of constant size, which allows the calculation of the distortion matrices as continuous forms depending of a unique angular parameter. The {225} habit plane of martensite in steels could be deduced from the KS OR by relaxing the IPS assumption of PTMC and by assuming that the interface plane is only untilted (Jaswon & Wheeler, 1948; Cayron, 2015). If required, the {225} unrotated plane can be rendered invariant (IPS) by coupling two KS variants in twin orientation relationship (Baur *et al*., 2017a). The exact KS OR imposes a strict invariant line along a dense direction that is common to fcc and bcc phase $[\bar{1}10]_\gamma$ // $[\bar{1}11]_\alpha$, which is fulfilled the hard-sphere assumption, but disagrees with the real lattice parameters used in PTMC. We have some reasons to believe that the continuous features observed in the pole figures martensite in steels come from the rotational incompatibilities between the low misorientated KS variants (Cayron, 2013), and that these incompatibilities are the consequence of the angular distortive character of the transformation. These rotational defects are disclinations; they play the same role in plasticity as dislocations for translational defects (Romanov, 2003); they are already used for phase transformations (see for example Müllner & King, 2010). Disclinations do not belong to the global group, at least with its present definition.

## Appendix B. Reminder on elementary matrices

### B1. Notations

To read the present paper, the reader should be familiar with basic linear algebra and group theory (subgroups, cosets, action of groups). If it is not the case, it can refer to Janovec, Hahn & Klapper (2003) or to the appendix of (Cayron, 2006). The vectors of a crystallographic basis are noted $\boldsymbol{a}$, $\boldsymbol{b}$, $\boldsymbol{c}$. The identity matrix is noted $\mathbf{I}$. The metric and structure tensors (matrices) are noted $\boldsymbol{\mathcal{M}}$ and $\boldsymbol{\mathcal{S}}$, respectively. In general in this paper the vectors are noted by bold lowercase letters and the matrices by bold uppercase letters, but there are some exceptions. A material point in continuum mechanics is usually noted $X$ (a vector made of the coordinates of the initial position of the point) in order to differentiate it from the trajectory of this point given by the spatial positions $x = \mathbf{F}.X$. The symmetry matrices are noted by the letter $g$ in italic lowercase to respect the usual notation in crystallography. Groups of matrices are noted by double-struck letters, for example $\mathbb{G}$. The letters are generally completed by superscripts and subscripts. The superscript refers to the phase (written in Greek letter) to which the vector or matrix belongs or refers. The subscript generally specifies the basis in which the vector or matrix is written. For example $\mathbf{u}^{\gamma}_{/\boldsymbol{\mathcal{B}}}$ means a direction $\mathbf{u}^{\gamma}$ of the phase γ whose coordinates are written in column in the crystallographic basis $\boldsymbol{\mathcal{B}}$. For sake of simplicity, the vector is simply noted $\mathbf{u}^{\gamma}_c$ when expressed in the crystallographic basis $\boldsymbol{\mathcal{B}}^{\gamma}_c = (\mathbf{a}^{\gamma}, \mathbf{b}^{\gamma}, \mathbf{c}^{\gamma})$, and $\mathbf{u}^{\gamma}_{\#}$ when it is expressed in the orthonormal basis linked to the crystallographic basis $\boldsymbol{\mathcal{B}}^{\gamma}_{\#}$ by the structure tensor $\boldsymbol{\mathcal{S}}^{\gamma}$. Note that distortion matrices are noted $\mathbf{F}^{\gamma}_c$ in this paper and not anymore $\mathbf{D}^{\gamma \rightarrow \alpha}_c$ as in (Cayron, 2015, 2016). The choice of the letter $\mathbf{F}$ was made to reinforce the link between crystallography and continuum mechanics, as it was already done by Battacharya (2003). The choice of replacing "γ→α" simply by "γ" in the superscript of $\mathbf{F}$ is more important. It was made to respect the head-tail composition of the superscripts, for example in the equation (5) $\mathbf{C}^{\alpha \rightarrow \gamma}_c = \mathbf{T}^{\alpha \rightarrow \gamma}_c \mathbf{F}^{\gamma}_c$, and to insist on the fact that the distortion alone cannot define the transformation, as shown in example of Figure 1. The correspondence or orientation should complement the distortion matrix to define the crystallographic characteristics of the transformation. The subscript is also sometimes used to indicate the index of an element in a group, for example $g^{\gamma}_i \in \mathbb{G}^{\gamma}$ means the $i^{\text{th}}$ element in the group of symmetries $\mathbb{G}^{\gamma}$ (this supposes that the matrices are enumerated). The inverse, transpose, and inverse of the transpose of a matrix are marked by the exponent -1, t, and *, respectively. The Greek letters γ and α are used for the phases, and θ or β are distortion angles. The notation $n/m$ with $n$ and $m$ integers means "$n$ divides $m$".

### B2. Coordinate transformation matrices and distortion matrices

The coordinate transformation matrix from the basis $\boldsymbol{\mathcal{B}}^{\gamma} = (\mathbf{a}^{\gamma}, \mathbf{b}^{\gamma}, \mathbf{c}^{\gamma})$ to the basis $\boldsymbol{\mathcal{B}}^{\alpha} = (\mathbf{a}^{\alpha}, \mathbf{b}^{\alpha}, \mathbf{c}^{\alpha})$ is $\mathbf{T}^{\gamma \rightarrow \alpha} = [\boldsymbol{\mathcal{B}}^{\gamma} \rightarrow \boldsymbol{\mathcal{B}}^{\alpha}]$; it is given by the coordinates of the vectors $\mathbf{a}^{\alpha}$, $\mathbf{b}^{\alpha}$, $\mathbf{c}^{\alpha}$ expressed in the crystallographic basis $\boldsymbol{\mathcal{B}}^{\gamma}$ and written in column. By explicitly writing $\mathbf{a}^{\alpha} = a^{\alpha}_{1/\gamma} \mathbf{a}^{\gamma} + a^{\alpha}_{2/\gamma} \mathbf{b}^{\gamma} +$ $a^{\alpha}_{3/\gamma} \mathbf{c}^{\gamma}$, $\mathbf{b}^{\alpha} = b^{\alpha}_{1/\gamma} \mathbf{a}^{\gamma} + b^{\alpha}_{1/\gamma} \mathbf{b}^{\gamma} + b^{\alpha}_{1/\gamma} \mathbf{c}^{\gamma}$ and $\mathbf{c}^{\alpha} = c^{\alpha}_{1/\gamma} \mathbf{a}^{\gamma} + c^{\alpha}_{1/\gamma} \mathbf{b}^{\gamma} + c^{\alpha}_{1/\gamma} \mathbf{c}^{\gamma}$; the linear relation can be written

$$(\mathbf{a}^{\alpha}, \mathbf{b}^{\alpha}, \mathbf{c}^{\alpha}) = (\mathbf{a}^{\gamma}, \mathbf{b}^{\gamma}, \mathbf{c}^{\gamma}) \cdot \mathbf{T}^{\gamma \to \alpha} \quad (B1)$$

with $\mathbf{T}^{\gamma \to \alpha} = \begin{bmatrix} a^{\alpha}_{1/\gamma} & b^{\alpha}_{1/\gamma} & c^{\alpha}_{1/\gamma} \\ a^{\alpha}_{2/\gamma} & b^{\alpha}_{2/\gamma} & c^{\alpha}_{2/\gamma} \\ a^{\alpha}_{3/\gamma} & b^{\alpha}_{3/\gamma} & c^{\alpha}_{3/\gamma} \end{bmatrix}$, or equivalently $\begin{bmatrix} \mathbf{a}^{\alpha} \\ \mathbf{b}^{\alpha} \\ \mathbf{c}^{\alpha} \end{bmatrix} = (\mathbf{T}^{\gamma \to \alpha})^{t} \begin{bmatrix} \mathbf{a}^{\gamma} \\ \mathbf{b}^{\gamma} \\ \mathbf{c}^{\gamma} \end{bmatrix}$.

A fixed vector $\mathbf{u}$ is expressed in the basis $\mathcal{B}^{\gamma}$ and in the basis $\mathcal{B}^{\alpha}$ by its coordinates written in column $\boldsymbol{u}_{/\gamma} = \begin{bmatrix} u_{1/\gamma} \\ u_{2/\gamma} \\ u_{3/\gamma} \end{bmatrix}$ and $\boldsymbol{u}_{/\alpha} = \begin{bmatrix} u_{1/\alpha} \\ u_{2/\alpha} \\ u_{3/\alpha} \end{bmatrix}$. The coordinates have a meaning only relatively to the basis in which they are written. The full expression of the vector is $\boldsymbol{u} = (\mathbf{a}^{\alpha}, \mathbf{b}^{\alpha}, \mathbf{c}^{\alpha}) \begin{bmatrix} u_{1/\alpha} \\ u_{2/\alpha} \\ u_{3/\alpha} \end{bmatrix} = (\mathbf{a}^{\gamma}, \mathbf{b}^{\gamma}, \mathbf{c}^{\gamma}) \begin{bmatrix} u_{1/\gamma} \\ u_{2/\gamma} \\ u_{3/\gamma} \end{bmatrix}$.

Combined with equation (B1), it implies that the matrix $\mathbf{T}^{\gamma \to \alpha}$ transforms the coordinates of the vectors according to $\begin{bmatrix} u_{1/\gamma} \\ u_{2/\gamma} \\ u_{3/\gamma} \end{bmatrix} = \mathbf{T}^{\gamma \to \alpha} \begin{bmatrix} u_{1/\alpha} \\ u_{2/\alpha} \\ u_{3/\alpha} \end{bmatrix}$, or equivalently, by noting the vectors formed by the column coordinates of the vector $\boldsymbol{u}$

$$\boldsymbol{u}_{/\gamma} = \mathbf{T}^{\gamma \to \alpha} \boldsymbol{u}_{/\alpha} \quad (B2)$$

When this relation is applied to the three basis vectors of the basis $\mathcal{B}^{\alpha}$, we write in series:

$$\left( \mathbf{a}^{\alpha}_{/\gamma}, \mathbf{b}^{\alpha}_{/\gamma}, \mathbf{b}^{\alpha}_{/\gamma} \right) = \mathbf{T}^{\gamma \to \alpha} \cdot \left( \mathbf{a}^{\alpha}_{/\alpha}, \mathbf{b}^{\alpha}_{/\alpha}, \mathbf{b}^{\alpha}_{/\alpha} \right) \quad (B3)$$

with obviously $\mathbf{a}^{\alpha}_{/\alpha} = \begin{bmatrix} 1 \\ 0 \\ 0 \end{bmatrix}$, $\mathbf{b}^{\alpha}_{/\alpha} = \begin{bmatrix} 0 \\ 1 \\ 0 \end{bmatrix}$ and $\mathbf{c}^{\alpha}_{/\alpha} = \begin{bmatrix} 0 \\ 0 \\ 1 \end{bmatrix}$. One can compare this equation with equation (B1). In equation (B1) $\mathbf{T}^{\gamma \to \alpha}$ establishes a relation between the triplets of vectors of two different bases, whereas in equation (B3) $\mathbf{T}^{\gamma \to \alpha}$ establishes a relation between the coordinates of a unique triplet of vectors written in two different bases.

It is also important to notice that the coordinate transformation matrices are passive matrices; they should be composed from the left to the right:

$$\mathbf{T}^{\gamma \to \beta} \mathbf{T}^{\beta \to \alpha} = \mathbf{T}^{\gamma \to \alpha} \quad (B4)$$

The coordinate transformation matrix of the inverse transformation $\mathbf{T}^{\gamma \to \alpha}$ is simply the inverse of the matrix of the direct transformations

$$\mathbf{T}^{\alpha \to \gamma} = (\mathbf{T}^{\gamma \to \alpha})^{-1} \quad (B5)$$

Now, let us consider a linear distortion $\mathbf{F}$ expressed by a matrix $\mathbf{F}_{/\gamma}$ in the basis $\mathcal{B}^{\gamma}$. A vector $\boldsymbol{u}$ expressed by its column vector $\boldsymbol{u}_{/\gamma}$ in the basis $\mathcal{B}^{\gamma}$ is transformed by the action $\mathbf{F}$ into a new vector $\boldsymbol{u}'$ expressed by its column vector $\boldsymbol{u'}_{/\gamma}$ in the basis $\mathcal{B}^\gamma$ such that $\boldsymbol{u'} = \mathbf{F}.\boldsymbol{u}$, expressed by the relation between the coordinates $\boldsymbol{u'}_{/\gamma} = \mathbf{F}_{/\gamma}.\boldsymbol{u}_{/\gamma}$. If a distortion $\mathbf{F}$ is combined with another distortion $\mathbf{G}$, the combination depends on the order of the actions. If $\mathbf{F}$ is applied first, $\boldsymbol{u'}_{/\gamma} = \mathbf{F}_{/\gamma}.\boldsymbol{u}_{/\gamma}$, and $\mathbf{G}^\gamma$ is applied in second, $\boldsymbol{u''}_{/\gamma} = \mathbf{G}_{/\gamma}.\boldsymbol{u'}_{/\gamma}$ then we get $\boldsymbol{u''}_{/\gamma} = (\mathbf{G}_{/\gamma}\mathbf{F}_{/\gamma}).\boldsymbol{u}_{/\gamma}$. The distortion matrices are active matrices; the distortion $\mathbf{F}$ followed by the distortion $\mathbf{G}$ is given by the matrix product $\mathbf{G}_{/\gamma}.\mathbf{F}_{/\gamma}$. The distortions should be composed from the right to the left when they are expressed in the same basis.

A distortion $\mathbf{F}$ expressed in $\mathcal{B}^\gamma$ by $\mathbf{F}_{/\gamma}$ and in $\mathcal{B}^\alpha$ by $\mathbf{F}_{/\alpha}$ is such that $\boldsymbol{u'}_{/\gamma} = \mathbf{F}_{/\gamma}.\boldsymbol{u}_{/\gamma}$ and $\boldsymbol{u'}_{/\alpha} = \mathbf{F}_{/\alpha}.\boldsymbol{u}_{/\alpha}$. By using the coordinate transformation matrix $\mathbf{T}^{\gamma \to \alpha} = [\mathcal{B}^\gamma \to \mathcal{B}^\alpha]$ and equation (B2), it comes immediately that

$$\mathbf{F}_{/\gamma} = \mathbf{T}^{\gamma \to \alpha}\,\mathbf{F}_{/\alpha}(\mathbf{T}^{\gamma \to \alpha})^{-1} \tag{B6}$$

**B3. Metric matrices**

A lattice is defined by its crystallographic basis $\mathcal{B}_c = (\mathbf{a}, \mathbf{b}, \mathbf{c})$. A vector $\boldsymbol{u}$ has coordinates that form the vector $\boldsymbol{u}_c = \begin{bmatrix} u_{1/c} \\ u_{2/c} \\ u_{3/c} \end{bmatrix}$. The vector is $\boldsymbol{u} = (\mathbf{a}, \mathbf{b}, \mathbf{c}) \begin{bmatrix} u_{1/c} \\ u_{2/c} \\ u_{3/c} \end{bmatrix}$. The scalar product of two vectors $\boldsymbol{u}$ and $\boldsymbol{v}$ is $\boldsymbol{u}^t.\boldsymbol{v} = (u_{1/c},\ u_{2/c},\ u_{3/c}) \begin{bmatrix} \mathbf{a}^t \\ \mathbf{b}^t \\ \mathbf{c}^t \end{bmatrix} (\mathbf{a}, \mathbf{b}, \mathbf{c}) \begin{bmatrix} v_{1/c} \\ v_{2/c} \\ v_{3/c} \end{bmatrix} = (u_{1/c},\ u_{2/c},\ u_{3/c})\,\mathcal{M} \begin{bmatrix} v_{1/c} \\ v_{2/c} \\ v_{3/c} \end{bmatrix} = \boldsymbol{u}^t_{/c}\,\mathcal{M}\,\boldsymbol{v}_{/c}$,

where

$$\mathcal{M} = \begin{bmatrix} \mathbf{a}^t \\ \mathbf{b}^t \\ \mathbf{c}^t \end{bmatrix} (\mathbf{a}, \mathbf{b}, \mathbf{c}) = \begin{bmatrix} \mathbf{a}^2 & \mathbf{b}^t\!\cdot\!\mathbf{a} & \mathbf{c}^t\!\cdot\!\mathbf{a} \\ \mathbf{a}^t\!\cdot\!\mathbf{b} & \mathbf{b}^2 & \mathbf{c}^t\!\cdot\!\mathbf{b} \\ \mathbf{a}^t\!\cdot\!\mathbf{c} & \mathbf{b}^t\!\cdot\!\mathbf{c} & \mathbf{c}^2 \end{bmatrix} \tag{B7}$$

The matrix $\mathcal{M}$ is the metric tensor; it contains the key parameters of the metrics of the lattice. Its components are given by the scalar products between the basis vectors, implicitly calculated by expressing these vectors into an orthornormal basis. The formula of scalar product between two crystallographic vectors helps to better understand why it is important to distinguish the vector from its coordinates:

$$\|\boldsymbol{u}\|^2 = \boldsymbol{u}^t.\boldsymbol{u} = \boldsymbol{u}^t_{/c}\,\mathcal{M}\,\boldsymbol{u}_{/c} \tag{B8}$$

From (B7) or (B8), it can be checked that the metric tensor is symmetric.

$$\mathcal{M} = \mathcal{M}^t \tag{B9}$$

The metric tensor can be also introduced by the using the reciprocal basis given by $\mathcal{B}_c^* = (\mathbf{a}^*, \mathbf{b}^*, \mathbf{c}^*)$ with $\mathbf{a}^* = \frac{\mathbf{b}\wedge\mathbf{c}}{V}$, $\mathbf{b}^* = \frac{\mathbf{c}\wedge\mathbf{a}}{V}$, $\mathbf{c}^* = \frac{\mathbf{a}\wedge\mathbf{b}}{V}$, with $V = det(\mathbf{a}\ \mathbf{b}\ \mathbf{c})$ the volume of the unit cell. This basis is such that

$$\begin{bmatrix} a \\ b \\ c \end{bmatrix} (a^*, b^*, c^*) = I \tag{B10}$$

By subtitling the vectors $(a, b, c)$ by $(a^*, b^*, c^*)$ in this equation and by inversing both sides, one gets that $(a^*)^*, (b^*)^*, (c^*)^* = a, b, c$, respectively, i.e $(\mathcal{B}_c^*)^* = \mathcal{B}_c$.

Let us consider a vector $u$ of coordinates in the direct and reciprocal bases $u_{/c} = \begin{bmatrix} u_{1/c} \\ u_{2/c} \\ u_{3/c} \end{bmatrix}$ and $u_{/*} = \begin{bmatrix} u_{1/*} \\ u_{2/*} \\ u_{3/*} \end{bmatrix}$. By using equation (B10), the norm of $u$ is simply $u^t . u = (u_{1/c}, u_{2/c}, u_{3/c}) \begin{bmatrix} u_{1/*} \\ u_{2/*} \\ u_{3/*} \end{bmatrix}$.

Comparing this expression with the previous one leads to $\begin{bmatrix} u_{1/*} \\ u_{2/*} \\ u_{3/*} \end{bmatrix} = \mathcal{M} \begin{bmatrix} u_{1/c} \\ u_{2/c} \\ u_{3/c} \end{bmatrix}$. Therefore, the metric tensor $\mathcal{M}$ can be understood as the coordinate transformation matrix from the reciprocal basis to the direct basis:

$$\mathcal{M} = [\mathcal{B}_c^* \rightarrow \mathcal{B}_c] \tag{B11}$$

With this equation, $\mathcal{M}^* = [(\mathcal{B}_c^*)^* \rightarrow \mathcal{B}_c^*] = [\mathcal{B}_c \rightarrow \mathcal{B}_c^*]$, which leads to

$$\mathcal{M}^* = \mathcal{M}^{-1} \tag{B12}$$

Instead of using the metric tensor, it is sometimes useful for the calculations to switch from the crystallographic basis to an orthonormal basis chosen according to an arbitrary rule. We call $\mathcal{B}_c = (a, b, c)$ the usual crystallographic basis, and $\mathcal{B}_\# = (x, y, z)$ the orthonormal basis linked to $\mathcal{B}_c$ by the structure tensor $\mathcal{S}$:

$$\mathcal{S} = [\mathcal{B}_\# \rightarrow \mathcal{B}_c] = \begin{pmatrix} a\sin(\alpha) & \dfrac{b(\cos(\gamma) - \cos(\alpha)\cos(\beta))}{\sin(\beta)} & 0 \\ 0 & \dfrac{bv}{\sin(\beta)} & 0 \\ a\cos(\beta) & b\cos(\alpha) & c \end{pmatrix} \tag{B13}$$

with $v = \dfrac{V}{abc} = \sqrt{1 + 2\cos(\alpha)\cos(\beta)\cos(\gamma) - \cos^2(\alpha) - \cos^2(\beta) - \cos^2(\gamma)}$

Any vector $u$ can be written in $\mathcal{B}_\#$ by $u_{/\#} = \begin{bmatrix} u_{1/\#} \\ u_{2/\#} \\ u_{3/\#} \end{bmatrix} = \mathcal{S} \begin{bmatrix} u_{1/c} \\ u_{2/c} \\ u_{3/c} \end{bmatrix}$, and its norm is simply $u^t . u = (u_{1/\#}, u_{2/\#}, u_{3/\#}) \begin{bmatrix} u_{1/\#} \\ u_{2/\#} \\ u_{3/\#} \end{bmatrix} = (u_{1/c}, u_{2/c}, u_{3/c}) \mathcal{S}^t \mathcal{S} \begin{bmatrix} u_{1/c} \\ u_{2/c} \\ u_{3/c} \end{bmatrix}$, which shows that

$$\mathcal{S}^t \mathcal{S} = \mathcal{M} \tag{B14}$$

Any rotation $Q$ is defined by an orthogonal matrix $Q_\#$ when written in an orthonormal basis $\mathcal{B}_\#$. As it preserves the scalar product,

$$Q_\#^t Q_\# = I \tag{B15}$$

When written in the crystallographic basis $\mathcal{B}_c$ the same rotation is written $Q_c = \mathcal{S}^{-1} Q_\# \mathcal{S}$. The preservation of the scalar product becomes

$$\mathbf{Q}_c^t \, \mathcal{M} \, \mathbf{Q}_c = \mathcal{M} \tag{B16}$$

which can be understood by the fact that the rotations preserve the metrics of the crystals.

**B4. Symmetry matrices**

One of the most fascinating properties of crystals is their symmetry. The shape formed by the faces of some macroscopic crystals in minerals perfectly reflects the symmetries of orientations of the planes. Let us consider the set of symmetries of crystallographic directions (point group) $\mathbb{G}$, which we will show later is the same as that of planes. This set is formed by the matrices that preserve the norms and angles of crystallographic directions, i.e. the scalar product defined with the crystal metrics. By default, the symmetry matrices will be always written in the crystallographic basis. With this simplification of notation, we write that any symmetry matrix $\boldsymbol{g}$ of the crystal is such that $\boldsymbol{g}^t \mathcal{M} \boldsymbol{g} = \mathcal{M}$ (see for example Rigault, 1980). In other words, the symmetries leave invariant the metrics $\mathcal{M}$ of the crystal; they are the stabilizer of the metrics by the conjugacy action. One can add the condition that they do not change the volume of the crystal, which imposes that the determinant of the matrix is $\pm 1$, i.e. the symmetry matrices belongs to the special linear group of dimension $N = 3$ on the real numbers, SL($N,\mathbb{R}$) or belongs to $-\mathbf{I}$ SL($N,\mathbb{R}$). A last constrain comes from the lattice: the symmetries should transform any vector of the crystallographic basis into another crystallographic vector, i.e. a vector of integer coordinates into another one, i.e. the matrix should be an integer matrix. These considerations can be combined to define the orientation symmetry matrices by

$$\mathbb{G} = \{\boldsymbol{g} \in \pm \text{SL}(N, \mathbb{Z}), \ \boldsymbol{g}^t \mathcal{M} \boldsymbol{g} = \mathcal{M}\} \tag{B17}$$

The group structure of $\mathbb{G}$ can be checked (remind that the inverse of a matrix is the transpose of the cofactor matrix times the inverse of its determinant).

Similarly, one can define $\mathbb{G}^*$ as the group of matrices of determinant $\pm 1$ and that leave the metric matrix $\mathcal{M}^*$ invariant: $\mathbb{G}^* = \{\boldsymbol{g} \in \pm \text{SL}(N, \mathbb{Z}), \ \boldsymbol{g}^t \mathcal{M}^* \boldsymbol{g} = \mathcal{M}^*\}$. With equations (B9) and (B12), it can be checked that $\mathbb{G}^* = \mathbb{G}$. One can say that the point group $\mathbb{G}$ is constituted by the symmetries that leave the metric invariant without specifying if the symmetries act on the directions or the planes.

**B5. Link between the metrics and symmetries of two phases**

The orientation relationship matrix $\mathbf{T}_c^{\gamma \to \alpha}$ that links the crystallographic bases of two crystals $\gamma$ and $\alpha$ can be used to establish some relations between the metric tensors of the two phases. We have defined in formula (B8) the metric tensor $\mathcal{M}$ by the matrix that allows calculating scalar product of vectors written in their crystallographic basis, $\boldsymbol{u}^t . \boldsymbol{v} = \boldsymbol{u}_{/c}^t \, \mathcal{M} \, \boldsymbol{v}_{/c}$. For two vectors $\boldsymbol{u}$ and $\boldsymbol{v}$, expressed simultaneously in the crystallographic basis $\mathcal{B}_c^\gamma$ of the crystal $\gamma$, and in the crystallographic basis $\mathcal{B}_c^\alpha$ of the crystal $\alpha$, it comes $\boldsymbol{u}^t . \boldsymbol{v} = \boldsymbol{u}_{/\gamma}^t \, \mathcal{M}^\gamma \, \boldsymbol{v}_{/\gamma} = \boldsymbol{u}_{/\alpha}^t \, \mathcal{M}^\alpha \, \boldsymbol{v}_{/\alpha}$. By using formula (B2) and the fact that the formula hold for any vectors $\boldsymbol{u}$ and $\boldsymbol{v}$, it comes that

$$\mathcal{M}^\gamma = \mathbf{T}_c^{\gamma\to\alpha} \mathcal{M}^\alpha \left(\mathbf{T}_c^{\gamma\to\alpha}\right)^{-1} \tag{B18}$$

The orientation relationship matrix $\mathbf{T}_c^{\gamma\to\alpha}$ also permits to establish some relations between the point groups of the two crystals. Let us apply the appendix formula (B17) that the group of symmetry matrices of the crystal γ and of crystal α,

$$\mathbb{G}^\gamma = \{\boldsymbol{g}^\gamma \in \pm\text{SL}(N,\mathbb{Z}),\ \boldsymbol{g}^{\gamma\text{t}}\mathcal{M}^\gamma\boldsymbol{g}^\gamma = \mathcal{M}^\gamma\} \tag{B19}$$
$$\mathbb{G}^\alpha = \{\boldsymbol{g}^\alpha \in \pm\text{SL}(N,\mathbb{Z}),\ \boldsymbol{g}^{\alpha\text{t}}\mathcal{M}^\alpha\boldsymbol{g}^\alpha = \mathcal{M}^\alpha\}$$

By using equation (B18) it can be shown that if $\boldsymbol{g}^\alpha \in \mathbb{G}^\alpha$, then $\mathbf{T}_c^{\gamma\to\alpha} \boldsymbol{g}^\alpha \left(\mathbf{T}_c^{\gamma\to\alpha}\right)^{-1}$ leaves invariant the metrics $\mathcal{M}^\gamma$. If the condition $\mathbf{T}_c^{\gamma\to\alpha} \boldsymbol{g}^\alpha \left(\mathbf{T}_c^{\gamma\to\alpha}\right)^{-1} \in \pm\text{SL}(N,\mathbb{Z})$ is fulfilled, then $\mathbf{T}_c^{\gamma\to\alpha} \boldsymbol{g}^\alpha \left(\mathbf{T}_c^{\gamma\to\alpha}\right)^{-1} \in \mathbb{G}^\gamma$. One can thus write

$$\mathbb{G}^\gamma \supset \mathbf{T}_c^{\gamma\to\alpha} \mathbb{G}^\alpha \left(\mathbf{T}_c^{\gamma\to\alpha}\right)^{-1} \cap \pm\text{SL}(N,\mathbb{Z}) \tag{B20}$$
$$\mathbb{G}^\alpha \supset \mathbf{T}_c^{\alpha\to\gamma} \mathbb{G}^\gamma \left(\mathbf{T}_c^{\alpha\to\gamma}\right)^{-1} \cap \pm\text{SL}(N,\mathbb{Z})$$

**Appendix C. Symmetry matrices for some 2D point groups**

Square P4mm (n°11), $\mathbb{G}^{Sq}$

$$g_1^{Sq} = \begin{pmatrix} 1 & 0 \\ 0 & 1 \end{pmatrix}, \ g_2^{Sq} = \begin{pmatrix} \bar{1} & 0 \\ 0 & 1 \end{pmatrix}, \ g_3^{Sq} = \begin{pmatrix} 1 & 0 \\ 0 & \bar{1} \end{pmatrix}, \ g_4^{Sq} = \begin{pmatrix} \bar{1} & 0 \\ 0 & \bar{1} \end{pmatrix},$$

$$g_5^{Sq} = \begin{pmatrix} 0 & 1 \\ 1 & 0 \end{pmatrix}, \ g_6^{Sq} = \begin{pmatrix} 0 & 1 \\ \bar{1} & 0 \end{pmatrix}, \ g_7^{Sq} = \begin{pmatrix} 0 & \bar{1} \\ 1 & 0 \end{pmatrix}, \ g_8^{Sq} = \begin{pmatrix} 0 & \bar{1} \\ \bar{1} & 0 \end{pmatrix}.$$

Rectangle P2mm (n°6), $\mathbb{G}^{Rc}$

$$g_1^{Rc} = \begin{pmatrix} 1 & 0 \\ 0 & 1 \end{pmatrix}, \ g_2^{Rc} = \begin{pmatrix} \bar{1} & 0 \\ 0 & 1 \end{pmatrix}, \ g_3^{Rc} = \begin{pmatrix} 1 & 0 \\ 0 & \bar{1} \end{pmatrix}, \ g_4^{Rc} = \begin{pmatrix} \bar{1} & 0 \\ 0 & \bar{1} \end{pmatrix}.$$

Hexagon P3m1 (n°14), $\mathbb{G}^{Tr}$ (called here Triangle)

$$g_1^{Tr} = \begin{pmatrix} 1 & 0 \\ 0 & 1 \end{pmatrix}, \ g_2^{Tr} = \begin{pmatrix} 0 & \bar{1} \\ 1 & \bar{1} \end{pmatrix}, \ g_3^{Tr} = \begin{pmatrix} \bar{1} & 1 \\ \bar{1} & 0 \end{pmatrix},$$

$$g_4^{Tr} = \begin{pmatrix} \bar{1} & 0 \\ 0 & \bar{1} \end{pmatrix}, \ g_5^{Tr} = \begin{pmatrix} \bar{1} & 1 \\ 0 & 1 \end{pmatrix}, \ g_6^{Tr} = \begin{pmatrix} 1 & 0 \\ 1 & \bar{1} \end{pmatrix}.$$

Hexagon P6mm (n°17), $\mathbb{G}^{Hx}$:

$$g_1^{Hx} = \begin{pmatrix} 1 & 0 \\ 0 & 1 \end{pmatrix}, \ g_2^{Hx} = \begin{pmatrix} 0 & \bar{1} \\ 1 & \bar{1} \end{pmatrix}, \ g_3^{Hx} = \begin{pmatrix} \bar{1} & 1 \\ \bar{1} & 0 \end{pmatrix},$$

$$g_4^{Hx} = \begin{pmatrix} \bar{1} & 0 \\ 0 & \bar{1} \end{pmatrix}, \ g_5^{Hx} = \begin{pmatrix} \bar{1} & 1 \\ 0 & 1 \end{pmatrix}, \ g_6^{Hx} = \begin{pmatrix} 1 & 0 \\ 1 & \bar{1} \end{pmatrix},$$

$$g_7^{Hx} = \begin{pmatrix} 0 & \bar{1} \\ \bar{1} & 0 \end{pmatrix}, \ g_8^{Hx} = \begin{pmatrix} 0 & 1 \\ \bar{1} & 1 \end{pmatrix}, \ g_9^{Hx} = \begin{pmatrix} 1 & \bar{1} \\ 1 & 0 \end{pmatrix},$$

$$g_{10}^{Hx} = \begin{pmatrix} 0 & 1 \\ 1 & 0 \end{pmatrix}, \ g_{11}^{Hx} = \begin{pmatrix} 1 & \bar{1} \\ 0 & \bar{1} \end{pmatrix}, \ g_{12}^{Hx} = \begin{pmatrix} \bar{1} & 0 \\ \bar{1} & 1 \end{pmatrix}.$$